 \documentclass[journal]{IEEEtran}

\usepackage{cite}

\ifCLASSINFOpdf
  \usepackage[pdftex]{graphicx}
  \graphicspath{{../pdf/}{../jpeg/}}
  \DeclareGraphicsExtensions{.pdf,.jpeg,.png}
\else
  \usepackage[dvips]{graphicx}
  \graphicspath{{../eps/}}
  \DeclareGraphicsExtensions{.eps}
\fi
\usepackage{amsmath}

\usepackage{array}

\ifCLASSOPTIONcompsoc
 \usepackage[caption=false,font=normalsize,labelfont=sf,textfont=sf]{subfig}
\else
 \usepackage[caption=false,font=footnotesize]{subfig}
\fi

\usepackage{dblfloatfix}

\usepackage{url}

\usepackage{amssymb}
\usepackage{amsthm}
\usepackage{bm}
\usepackage{mathtools}
\usepackage{dsfont}

\usepackage{url}
\usepackage{color,soul}

\usepackage[percent]{overpic}
\usepackage{siunitx}
\newcommand{\op}{\operatorname}
\usepackage{float}

\IEEEpubid{\begin{minipage}{\textwidth}\ \\[12pt]
    Copyright (c) 2019 IEEE. Personal use of this material is permitted. 
    However, permission to use this material for any other purposes must be
    obtained from the IEEE by sending a request to pubs-permissions@ieee.org
\end{minipage}} 

\begin{document}

\title{Spatio-Temporal Deep Learning-Based Undersampling Artefact Reduction for $2D$ Radial Cine MRI with Limited Training Data}

\author{Andreas~Kofler,
        Marc~Dewey,
        Tobias~Schaeffter,
        Christian~Wald,
        and~Christoph~Kolbitsch
\thanks{A. Kofler and C. Wald are with the Department of Radiology, Charit\'{e} - Universit\"{a}tsmedizin Berlin, Berlin,
Germany (e-mail: $\{$andreas.kofler, christian.wald$\}$@charite.de)}
\thanks{M. Dewey is with the  Department of Radiology, Charit\'{e} - Universit\"{a}tsmedizin Berlin, Berlin,
Germany and the Berlin Institute of Health, Berlin, Germany (e-mail: $\{$marc.dewey$\}$@charite.de)}
\thanks{T. Schaeffter and C. Kolbitsch are with the Physikalisch-Technische Bundesanstalt (PTB), Braunschweig and Berlin, Germany and King’s College London, London, UK
(e-mail:$\{$tobias.schaeffter, christoph.kolbitsch$\}$@ptb.de)}
}

\markboth{Accepted for Publication in IEEE Transactions on Medical Imaging}
{Shell \MakeLowercase{\textit{et al.}}: Bare Demo of IEEEtran.cls for IEEE Journals}

\maketitle

\begin{abstract}
In this work we reduce undersampling artefacts in two-dimensional ($2D$) golden-angle radial cine cardiac MRI by applying a modified version of the U-net. The network is trained on  $2D$ spatio-temporal slices which are previously extracted from the image sequences. We compare our approach to two $2D$ and a $3D$ Deep Learning-based post processing methods, three iterative reconstruction methods and two recently proposed methods for dynamic cardiac MRI based on $2D$ and $3D$ cascaded networks.  Our method outperforms the $2D$ spatially trained U-net and the $2D$ spatio-temporal U-net. Compared to the $3D$ spatio-temporal U-net, our method delivers comparable results, but requiring shorter training times and less training data. Compared to the Compressed Sensing-based methods $kt$-FOCUSS and a total variation regularized reconstruction approach, our method improves image quality with respect to all reported metrics. Further, it achieves competitive results when compared to the iterative reconstruction method based on adaptive regularization with Dictionary Learning and total variation and when compared to the methods based on cascaded networks, while only requiring a small fraction of the computational and training time. 
A persistent homology analysis demonstrates that the data manifold of the spatio-temporal domain has a lower complexity than the one of the spatial domain and therefore, the learning of a projection-like mapping is facilitated. Even when trained on only one single subject without data-augmentation, our approach yields results which are similar to the ones obtained on a large training dataset. This makes the method particularly suitable for training a network on limited training data.  Finally, in contrast to the spatial $2D$ U-net, our proposed method is shown to be naturally robust with respect to image rotation in image space and almost achieves rotation-equivariance where neither data-augmentation nor a particular network design are required.

\end{abstract}


\IEEEpeerreviewmaketitle

\section{Introduction}

\IEEEPARstart{M}{agnetic} Resonance Imaging (MRI) is a widely used non-invasive imaging modality in clinical practice. Especially for cardiac applications, MRI does not only provide anatomical imaging with excellent soft tissue contrast but also allows for functional assessment by using 2D cine MRI. Such images show the heart anatomy for different phases of the cardiac cycle providing valuable information of the heart function  \cite{Kramer2013, VonKnobelsdorff-Brenkenhoff2017}.

However, MRI suffers from long data-acquisition which determines the achievable spatial and temporal resolution. In order to shorten scan times, ensure sufficiently large spatial coverage and high spatial and temporal resolution, a wide range of undersampling and reconstruction techniques have been proposed, ranging from Parallel Imaging to Compressed Sensing (CS) and Dictionary Learning \cite{Tsao2003} - \cite{block2007undersampled}.
Cine MRI provides a temporal sequence of images and therefore offers the possibility to exploit the temporal correlation of adjacent frames in order to reduce undersampling artefacts. The movement of the heart during the cardiac cycle is mainly smooth and continuous. Ensuring that undersampling artefacts along time are incoherent and using a sparsifying transform along time such as Fourier transform \cite{Tsao2003}, Principal Component Analysis \cite{Pedersen2009, gupta2001dynamic}, Wavelet transform \cite{Jung2009} or a transform learned from data \cite{caballero2014dictionary,wang2014compressed} combined with a $L_1$-norm minimization approach can strongly reduce undersampling artefacts. The main challenges of these techniques are to ensure that the sparsifying transform really leads to a sparse signal and long reconstruction times due to the iterative reconstruction. 

Recently, Neural Networks (NNs) have been applied to inverse problems as image reconstruction in MRI \cite{jin2017deep}, \cite{sun2016deep},  \cite{hammernik2018learning}, \cite{zhu2018image} and computed tomography (CT) \cite{jin2017deep}, \cite{han2016deep}, \cite{kang2017deep}. Autoencoders, and in particular the U-net \cite{ronneberger2015u}, a convolutional NN (CNN)  which was first introduced for biomedical image segmentation, and different derivations of it \cite{ye2018deep}, \cite{han2018framing}, have been widely used for removing undersampling artefacts in different medical imaging modalities. \\ In initial works, the images were most commonly reconstructed or processed frame by frame, see e.g.  \cite{jin2017deep}. In the case of dynamic MRI, however, the temporal correlation of $2D$ MRI sequences can be exploited by aligning frames along the channel axis. Thus, $2D$ CNNs can be trained to map whole undersampled image sequences to their corresponding fully sampled image sequences \cite{schlemper2017deep,sandino2017deep}. Further, also CNNs employing $3D$-convolutions were shown to be trainable on entire image sequences, either as post-processing methods \cite{Hauptmann2019} or as unrolled iterative reconstruction schemes \cite{schlemper2017deep}. However, in general, due to the resulting high dimensionality of the considered problem, either a large dataset or the application of data-augmentation techniques are indispensable to obtain satisfactory results, see e.g. \cite{schlemper2017deep, Hauptmann2019}.\\

Nowadays it is common practice to learn the filters of the convolutional layers by considering the images in the spatial domain. In this work, we propose to apply CNNs to two-dimensional slices extracted from the spatio-temporal dimension in order to remove undersampling artefacts from a $2D$ cine MR scan obtained with a $2D$ Golden radial sampling scheme \cite{kolbitsch_mrm_2013}. A persistent homology analysis shows that the manifold of the spatio-temporal slices has a lower topological complexity than the manifold of the two-dimensional spatial image frames and suggests that the learning process of the network can therefore be facilitated. We compare our proposed approach to a $2D$ U-net trained frame-by-frame \cite{jin2017deep}, a $2D$ U-net trained image sequence-wise \cite{sandino2017deep} and a $3D$ U-net \cite{Hauptmann2019} in terms of image quality, amount of required training data and stability with respect to rotation of the images. The latter is important because $2D$ cine MRI is commonly obtained in oblique planes which are adapted to the patients anatomy. Our spatio-temporal approach method is also compared to three CS-based approaches for image reconstruction of cine MRI: $kt$-FOCUSS \cite{jung_magn_reson_med_01_2010}, a total variation minimization-based method \cite{block2007undersampled} and a Dictionary Learning- and total variation-based reconstruction method \cite{wang2014compressed}. Further, we compare our method to two methods for cine MRI based on cascaded networks \cite{schlemper2017deep}, \cite{qin2019convolutional}.

The paper is organized as follows. In Section \ref{section_Problem_Formulation}, we shortly discuss how NNs have been integrated within the problem of image reconstruction in MRI so far. Section \ref{section_Proposed_Approach} introduces our proposed method by discussing an a priori performed persistent homology analysis of the data which is needed to derive the approach as well as the network's architecture. We then show results of In-Vivo experiments and compare our method to other Deep Learning- and CS-based methods in Section \ref{section_In_Vivo_Experiments} and finish with a discussion and conclusion in Section \ref{section_Discussion_and_Conclusion}.

\section{Problem Formulation}\label{section_Problem_Formulation}

In dynamic MRI, the image reconstruction problem is given by finding a solution of the inverse problem 
\begin{equation}
\mathbf{F}\mathbf{x} = \mathbf{y},
\end{equation}
where $\mathbf{x}\in \mathbb{C}^N$ denotes the complex-valued image sequence with $N=N_x N_y N_t$, $\mathbf{F}$ denotes the Fourier encoding matrix and $\mathbf{y}$ corresponds to the measured data in $k$-space. As the data-acquisition process in MRI is slow, undersampling schemes are applied to fasten the measurement process. Therefore, the inverse problem one encounters in applications is of the form
\begin{equation}\label{sparse_problem}
\mathbf{F}_{I}\mathbf{x} = \mathbf{y}_I,
\end{equation}
where $\mathbf{F}_I = \mathbf{S}_I \mathbf{F}$ and $\mathbf{S}_I \in \mathbb{C}^{M\times N}$ denotes a binary undersampling operator with $M\ll N$ which sets non-measured values in $k$-space to zero. Thereby, $I \subset J = \{1,\ldots,N\}$ corresponds to the set of indices of the measured Fourier coefficients. Since $M\ll N$, the problem in (\ref{sparse_problem}) is underdetermined and therefore ill-posed. Hence, a direct solution is not possible and usually regularization approaches have to be applied in order to constrain the sought solution. Two widely used regularization techniques are based on Dictionary Learning \cite{caballero2014dictionary, wang2014compressed} and total-variation (TV) minimization \cite{block2007undersampled, wang2008new}. However, since the methods employ the regularization within an iterative reconstruction, solving the problem in (\ref{sparse_problem}) is time consuming and NNs have been considered as a valid and powerful alternative, see e.g.\ \cite{jin2017deep},  \cite{sun2016deep}, \cite{hammernik2018learning}, \cite{schlemper2017deep}, \cite{Hauptmann2019}. 

Most commonly, the networks are trained by considering the images in the spatial domain.   By doing so, the network learns to distinguish between diagnostic content of the image and the artefacts by exploiting the natural correlation of neighbouring pixel values in spatial domain. Given a dynamic process, one can further make use of the correlation of temporal slices amongst each other. 
In \cite{sandino2017deep}, the work of \cite{jin2017deep} is extended in the sense that a U-net is trained to directly map whole $2D$ image sequences of undersampled image reconstructions to $2D$ image sequences of ground truth images. In \cite{schlemper2017deep}, the temporal dimension of the sequence is taken into account in the same manner, where furthermore, a weighted data-sharing and a data-consistency approach further improve the quality of the reconstruction. For the $2D$ networks, frames corresponding to different cardiac phases are aligned along the channel axis.  As shown in \cite{schlemper2017deep} and \cite{Hauptmann2019}, CNNs employing $3D$ convolutional layers can also be applied for the task of removing undersampling artefacts in dynamic sequences.
Note that, for a network employing $2D$ convolutional layers and assuming the channel's dimension to be the one along which feature maps are combined by linear combination, aligning temporal frames along the channel's axis only slightly increases the  computational complexity of the CNN. In this case, the filters size only increases for the first and the last convolutional layers. Employing $3D$ convolutional layers, in contrast, adds further non-negligible computational cost as well as hardware requirements, increases training time, the number of trainable parameters and therefore the number of samples required to successfully train a network without experiencing overfitting.\\
In the aforementioned methods, the resulting number of available training samples reduces to the number of $2D$ image sequences. Since NNs are well known to require a large number of training samples and as the collection of proper data can be challenging, using these approaches, one usually has to heavily rely on the use of data-augmentation techniques, see e.g. \cite{schlemper2017deep}, have access to a large dataset \cite{Hauptmann2019} or both in order to obtain a good representation of the data manifold.
However, data-augmentation might also be non-trivial, time consuming or not possible to be performed on the fly. In the case of image reconstruction, the dataset is obtained by a prior data-acquisition process. In a simulation-based framework, one can for example apply arbitrary transformations to a ground truth image, e.g. elastic transformations, and then simulate the data-acquisition process. Also, using different undersampling masks to obtain zero-filled reconstructions can further enrich the data, see for example \cite{schlemper2017deep, sandino2017deep}. However, assuming a fixed dataset of pairs of undersampled image reconstructions and ground truth images, transformations would have to be applied to each pair, possibly altering the structure of the undersampling artefacts in the input images. \\ The same holds true for including rotated versions of training pairs into the dataset. As CNNs are not  necessarily rotation-invariant or rotation-equivariant,  these properties are usually achieved by properly augmenting the dataset \cite{krizhevsky2012imagenet}. In contrast, other approaches explicitly incorporate mathematical operations in the design of the network architectures and therewith attempt to reach rotation-invariance or -equivariance \cite{worrall2017harmonic}, \cite{marcos2017rotation}. High quality images in cardiac MRI are usually reconstructed by applying iterative methods. Thus, obtaining realistic versions of images rotated by a non-trivial rotation, i.e. by a rotation of $\theta \not \in \{\frac{k \pi}{2}: k \in \{0,1,2,3\}\}$, is computationally demanding, as the $k$-space data has to be rotated and the iterative reconstruction has to be performed on the rotated data. Therefore, rotation-equivariance, in this case, can either be achieved by means of the network architecture design or by a possibly time consuming data-augmentation process.

\section{Proposed Approach}\label{section_Proposed_Approach}

In medical imaging, the number of available training samples is usually very small compared to the underlying mathematical dimension of the data, i.e. the number of pixels of an image. Therefore,  we are particularly interested in the question of whether or not it is possible to train a CNN on a highly limited dataset by making best use of the given data.  We propose to train a  CNN employing $2D$ convolutional layers on $2D$ spatio-temporal slices which can be extracted from the cine image sequences over the cardiac cycle. Once the network is trained, the image sequences can be reconstructed by properly reassembling the spatio-temporal slices. Later, we demonstrate that with our proposed approach, already a small number of $2D$ cine MRI datasets suffices to successfully train a network. Furthermore, robustness with respect to rotation in the spatial domain is achieved in a natural way by the change of perspective on the given dataset and our method is therefore almost rotation-equivariant.

Consider a dataset of $2D$ cine MR images $\mathcal{D}$ of $n$ subjects, each with $N_z$ slices of size $N_x \times N_y$ and $N_t$ cardiac phases. Figure \ref{different_approaches} shows different possible Deep Learning-based methods for removing undersampling artefacts in dynamic MRI sequences. In the first case, the artefacts are removed by training a network $f_{\Theta}$ to map frames to frames, see Figure \ref{different_approaches} (a). Given the temporal correlation of adjacent frames, one could also align temporal frames along the channel's axis and apply a network which is trained to map whole image sequences to image sequences, see Figure \ref{different_approaches} (b). The same approach can be extended to map image sequences to image sequences but with the network employing three-dimensional convolutional filters, see Figure \ref{different_approaches} (c). Our approach exploits spatio-temporal correlation but employs $2D$ convolutional filters which are trained on the spatio-temporal slices of the image sequences, see Figure \ref{different_approaches} (d).  Table \ref{table_different_approaches}  lists the number of immediately available training samples, i.e.\ without data augmentation, for the different approaches.  Note that with our proposed approach, the number is by far the highest.
\begin{figure}[!h]
\centering
\begin{overpic}[width=\linewidth,tics=10]{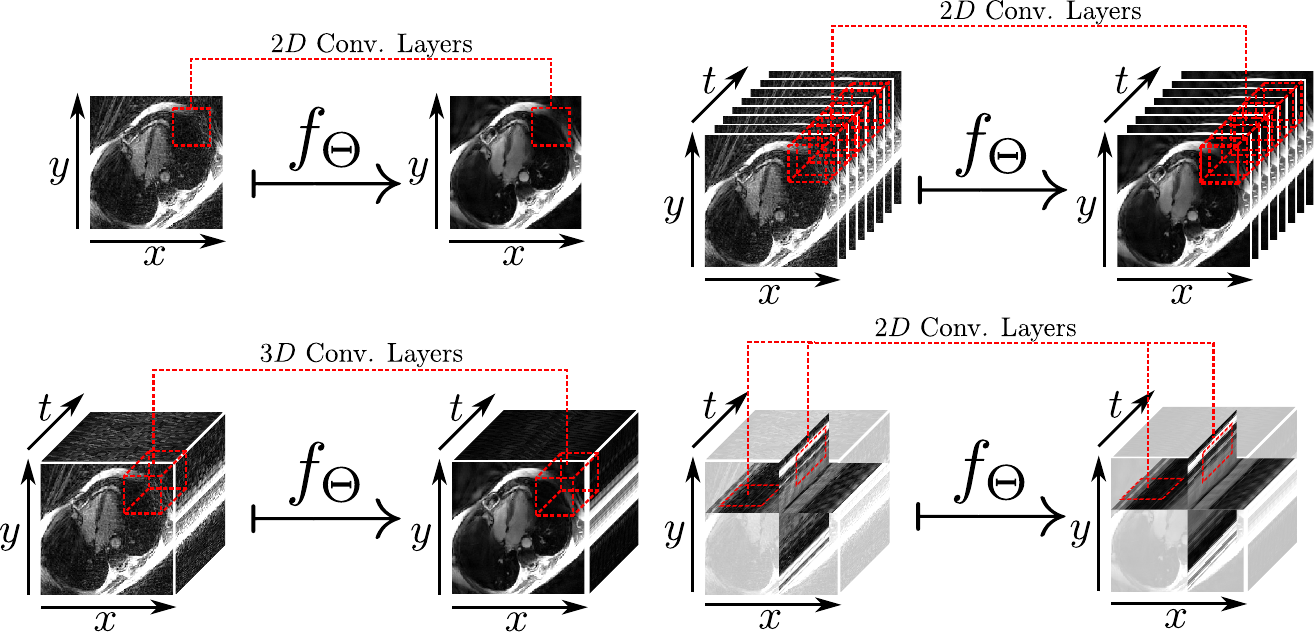}
 \put (23,27) {\small{(a)}}
 \put (73,27) {\small{(b)}}
\put (23,2) {\small{(c)}}
\put (73,2) {\small{(d)}}
\end{overpic}
\caption{Different $2D$ and $3D$ Deep Learning-based approaches for undersampling artefacts reduction. $2D$ network for frame-wise mapping (a), $2D$ network for image sequence-wise mapping with cardiac phases aligned as channels (b), $3D$ network for image sequence-wise mapping with three-dimensional convolutional kernels (c), $2D$ network for our proposed approach on two-dimensional spatio-temporal slices (d).}
\label{different_approaches}
\end{figure}
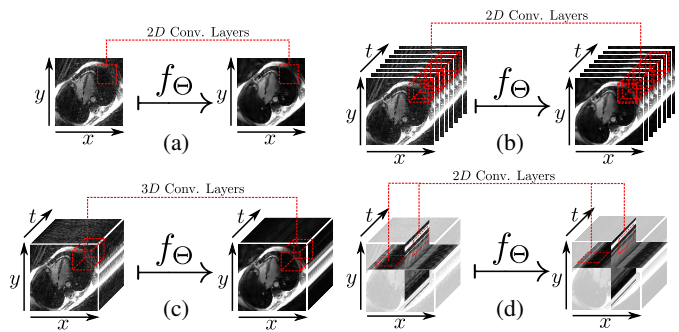 
\begin{table}[!h]
\renewcommand{\arraystretch}{1.3}
\begin{center}
\caption{Different Deep Learning-based approaches with their corresponding number of available training samples.}\label{table_different_approaches}
  \begin{tabular}{lll}
   \hline
   \textbf{Approach} & \textbf{Conv. Layers} & \textbf{Available Training Samples}\\
   \hline
    Frame-wise & $2D$ & $n \cdot N_z\cdot N_t$\\
    Sequence-wise & $2D$ & $n \cdot N_z$\\
    Sequence-wise & $3D$ & $n \cdot N_z$\\
    Proposed & $2D$ & $n \cdot (N_x+N_y)\cdot N_z$\\
    \hline
  \end{tabular}  
\end{center}
\end{table}
\subsection{Persistent Homology Analysis}\label{pers_hom_sec}

\begin{figure}[!h]
\centering
\begin{overpic}[width=0.85\linewidth,tics=10]{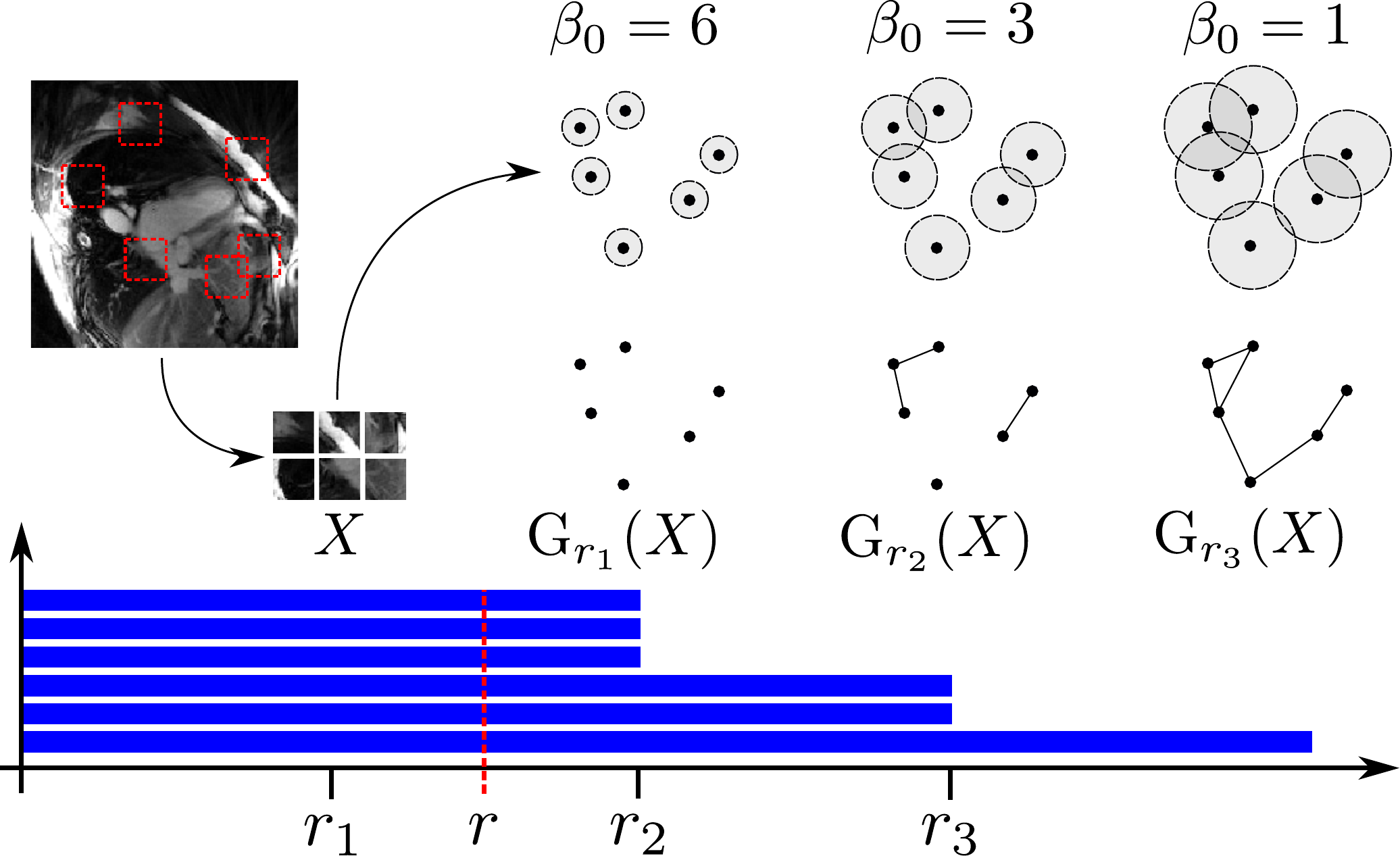}
 \end{overpic}
\caption{Procedure of the persistent homology analysis. The image shows an example for six randomly extracted patches of an image in the spatial domain and its corresponding barcode.}\label{connected_components}
\end{figure}

As a trained denoising autoencoder can  geometrically be interpreted to perform a projection-like mapping onto a manifold \cite{vincent2010stacked}, the study of topological features of the manifold of the input and output images might be of  interest for the design of the network architecture, \cite{han2016deep}, \cite{bae2017beyond}.
Persistent homology is a mathematical tool that can be used for analysing datasets  $X \subset \mathbb{R}^n$ \cite{ghrist}. For a two-classes classification problem, singular homology has been used as a complexity measure of the positively labelled submanifold of the input space and a relation between this complexity and the depth of the networks was proven in \cite{bianchini2014complexity}. This and experimental evidence using persistent homology \cite{han2016deep, bae2017beyond}, motivates that it might be beneficial to investigate the persistent homology of datasets since it might explain the superiority of specific approaches to others.
For a concise introduction to persistent homology see \cite{oudot2015persistence}, Chapter 1. In general, persistent homology $H_{*}$ assigns a family of persistence modules 
$\left\{H_{i}(X):i\in\mathbb{N}\right\}$ over some field $F$ to a set $X\subset\mathbb{R}^n$, see \cite{oudot2015persistence}, Chapter $2$. We will only use $H_{0}$ which has a much simpler interpretation as follows, see Figure \ref{connected_components}. Let $X\subset \mathbb{R}^n$ be a finite set and let $r\geq0$. Then, we can define a graph $\operatorname{G}_r(X)$ with vertices $\operatorname{V}_r(X)=X$ and edges 
\[
\operatorname{E}_r(X)=\left\{(x,y)\in X^2:x\neq y \text{ and }\|x-y\|_2\leq r\right\}.
\]
This graph is the Rips complex restricted to simplices of dimension at most $1$ \cite{ghrist}, Chapter 1.3. Let $\Pi(\op{G}_r(X))$ be the set of connected components of $\op{G}_r(X)$. Then, we can define \[H_0^r= {\bigoplus}_{i \in \Pi(\op{G}_r(X))} \mathbb{F}_2\] where $\mathbb{F}_2$ is the field with two elements. For $0\leq r<r'$ we have a map $\Pi(\op{G}_r(\op{X}))\rightarrow\Pi(\op{G}_{r'}(\op{X}))$ which induces a map $H_0^r\rightarrow H_0^{r'}.$ The family of these maps is called the $0$-th persistent homology of $X$. A good visualization of persistent homology is the persistent barcode, see Figure \ref{connected_components}. For a real number $r>0$, the number of connected components of $\op{G}_r(X)$ is equal the number of intersections of the vertical line at $x=r$ with the barcodes, see Figure \ref{connected_components}. This is also the $0$-th Betti number $\beta_0$ of $\op{G}_r(X)$ which is a measure of complexity for $\op{G}_r(X)$, see \cite{ghrist},  Chapter 2.3. Hence, the faster the persistent barcode of a dataset $X$ decreases, the less complex the dataset is.

By $\mathbf{x}_I, \mathbf{x}$ and $\mathbf{r}_I:=\mathbf{x}_I-\mathbf{x}$ we denote the vector representations of direct reconstruction from undersampled radially acquired data using a non-uniform fast Fourier transform approach (NUFFT), the ground truth reconstruction and the residual, respectively. Since our network reduces artefacts arising from the NUFFT reconstruction as a post-processing step similar to denoising, we operate on the real-valued magnitude images. However, the method can also be applied to complex-valued images by treating real- and imaginary part separately. Note that, in order to keep notation as simple as possible, by abuse of notation, we do not explicitly distinguish between a spatio-temporal slice and a $2D$ frame, but the meaning of the symbols should easily emerge from the context. Therefore, in the spatio-temporal training scenario, $\mathbf{x}_I$ denotes a spatio-temporal slice extracted from an undersampled NUFFT reconstruction, $\mathbf{x}$ its corresponding artefact-free spatio-temporal slice and $\mathbf{r}_I$ its spatio-temporal residual. In the spatial training scenario, $\mathbf{x}_I$, $\mathbf{x}$ and $\mathbf{r}_I$ denote $2D$ frames. 
In the following, we compare the complexity of the manifolds given by the set of the ground truth images and their residuals in the spatial as well as in the spatio-temporal domain and denote them by $\mathcal{M}_{xy}^{\mathrm{img}}$, $\mathcal{M}_{xy}^{\mathrm{res}}$ and $\mathcal{M}_{xt,yt}^{\mathrm{img}}$, $\mathcal{M}_{xt,yt}^{\mathrm{res}}$. Note that, in contrast to \cite{han2016deep}, we find ourselves in the situation where spatio-temporal slices and spatial images do not have the same mathematical dimension, and therefore, to be able to compare the manifolds, we restrict our considerations to image patches of the same shape.
\begin{figure}[h]
\begin{overpic}[width=0.482\linewidth,tics=10]{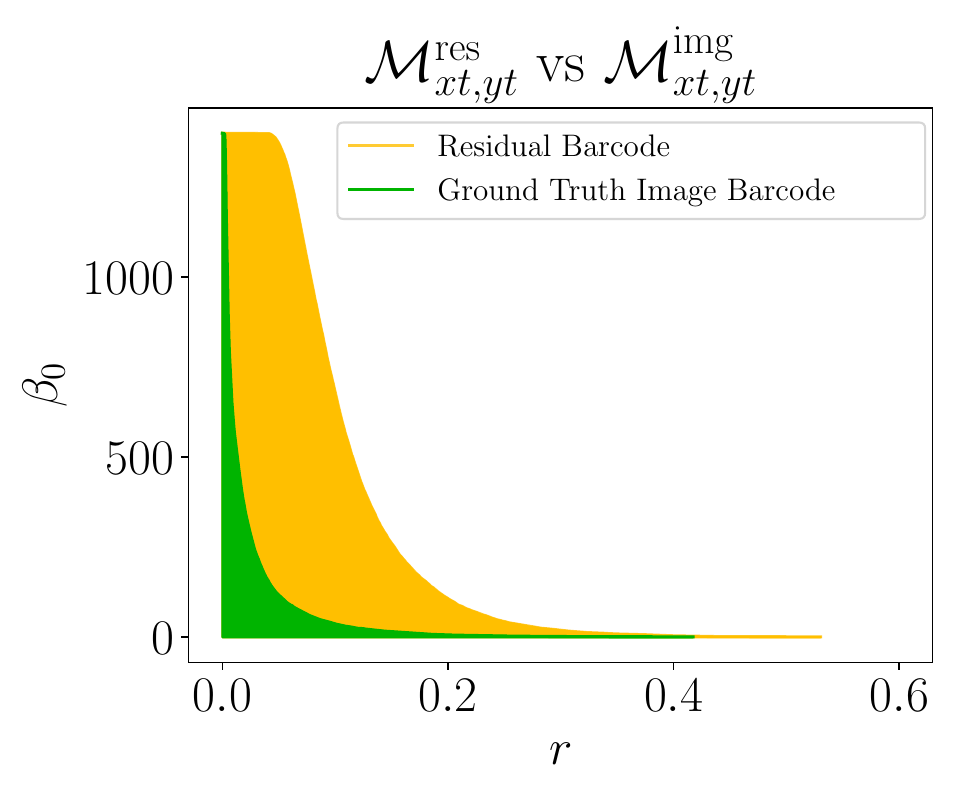}
\put (80,20) {\large{(a)}}
\end{overpic}
\begin{overpic}[width=0.482\linewidth,tics=10]{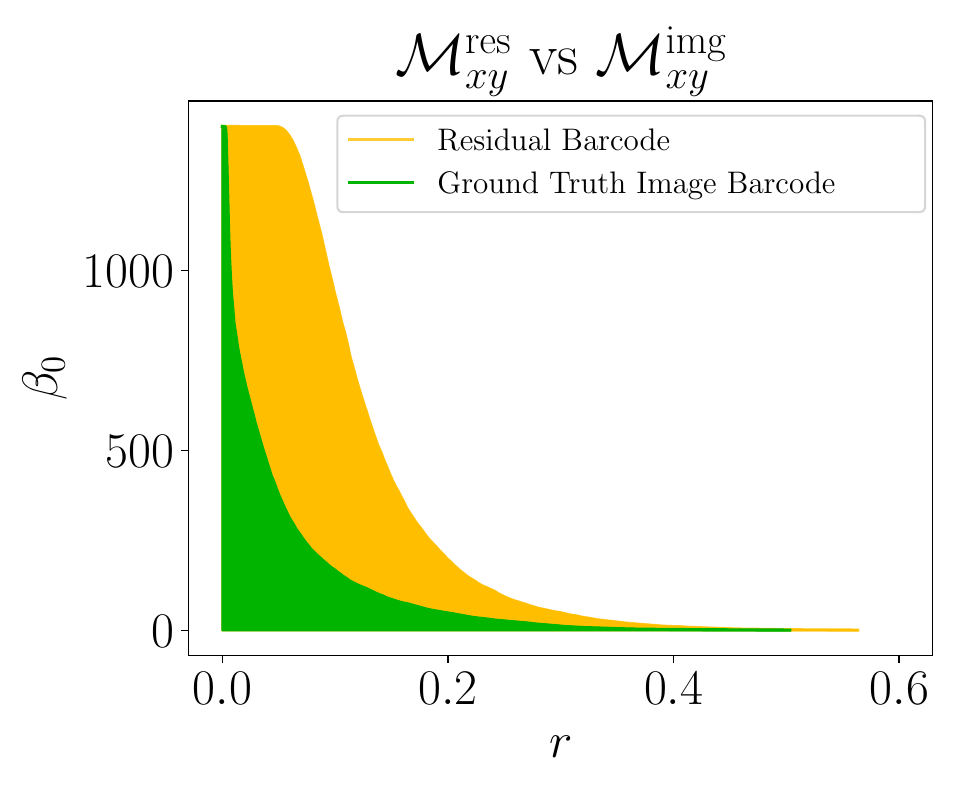}
\put (80,20) {\large{(b)}}
\end{overpic}
\begin{overpic}[width=0.482\linewidth,tics=10]{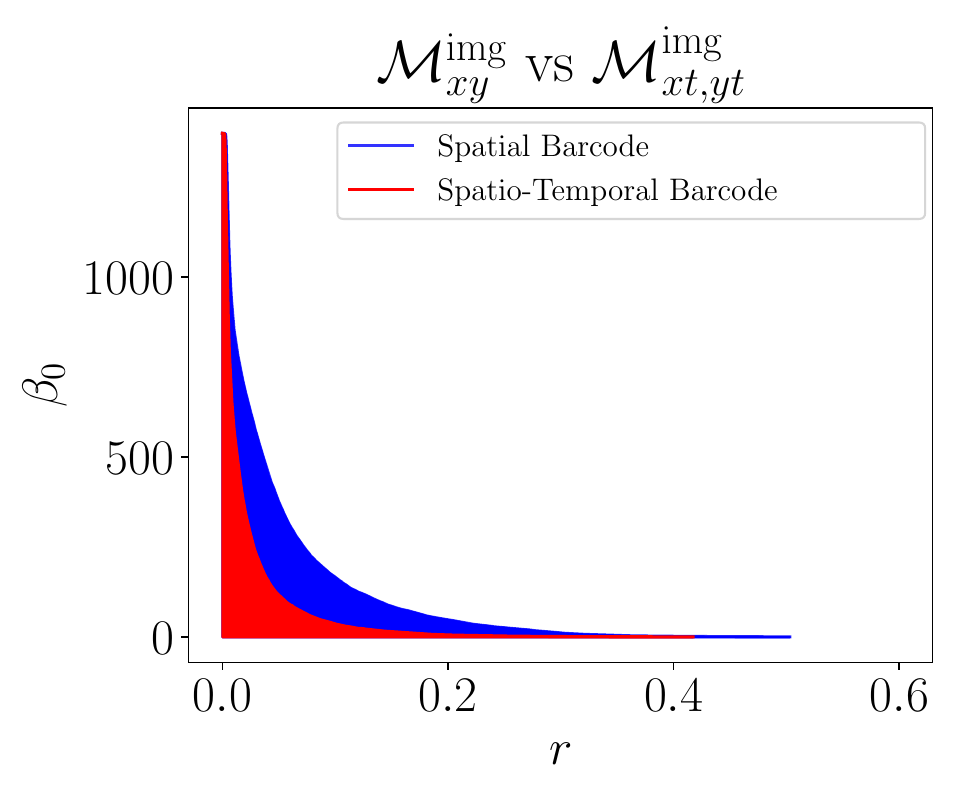}
\put (80,20) {\large{(c)}}
\end{overpic}
\begin{overpic}[width=0.482\linewidth,tics=10]{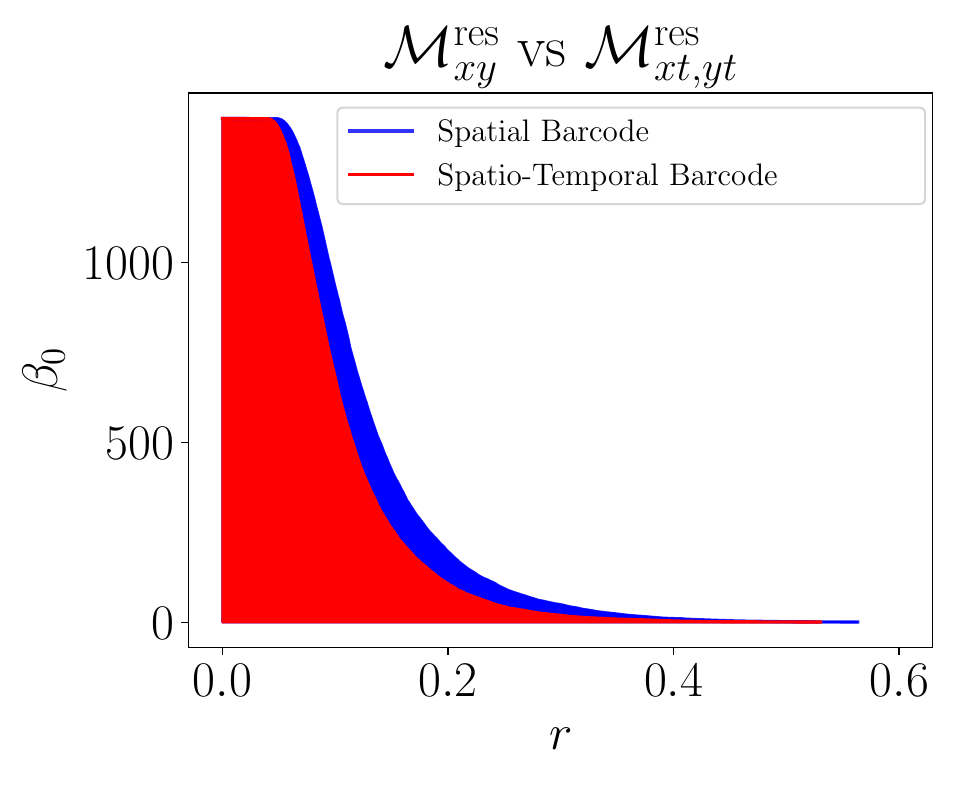}
\put (80,20) {\large{(d)}}
\end{overpic}
\begin{overpic}[width=0.482\linewidth,tics=10]{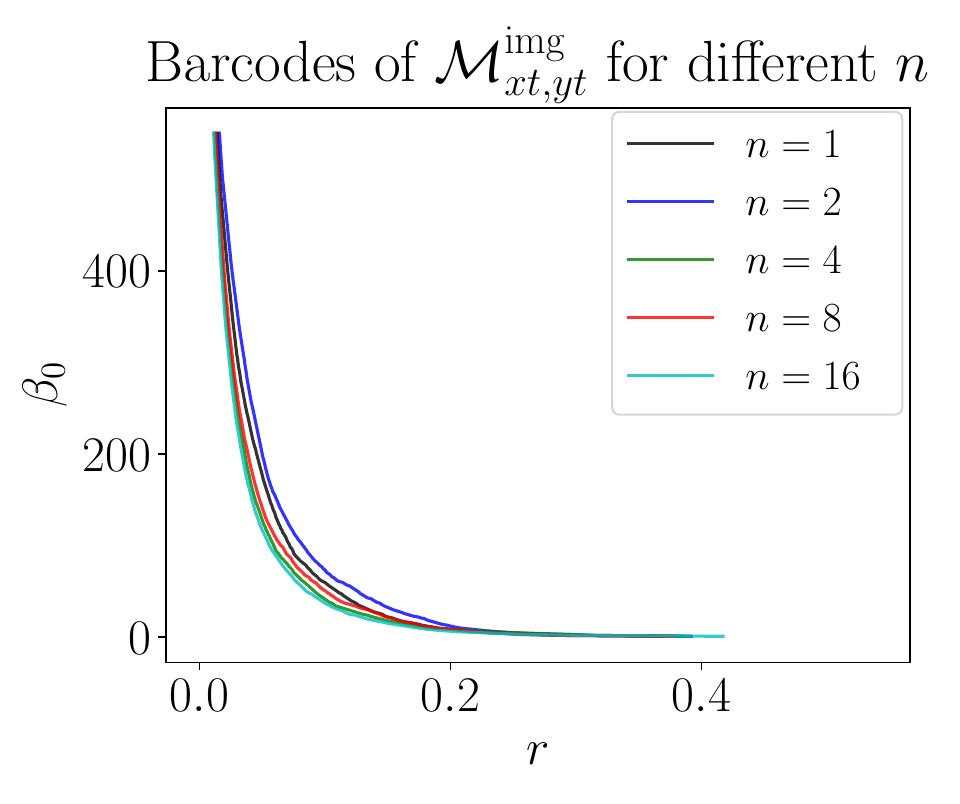}
\put (80,20) {\large{(e)}}
\end{overpic}\hspace{0.3cm}
\begin{overpic}[width=0.482\linewidth,tics=10]{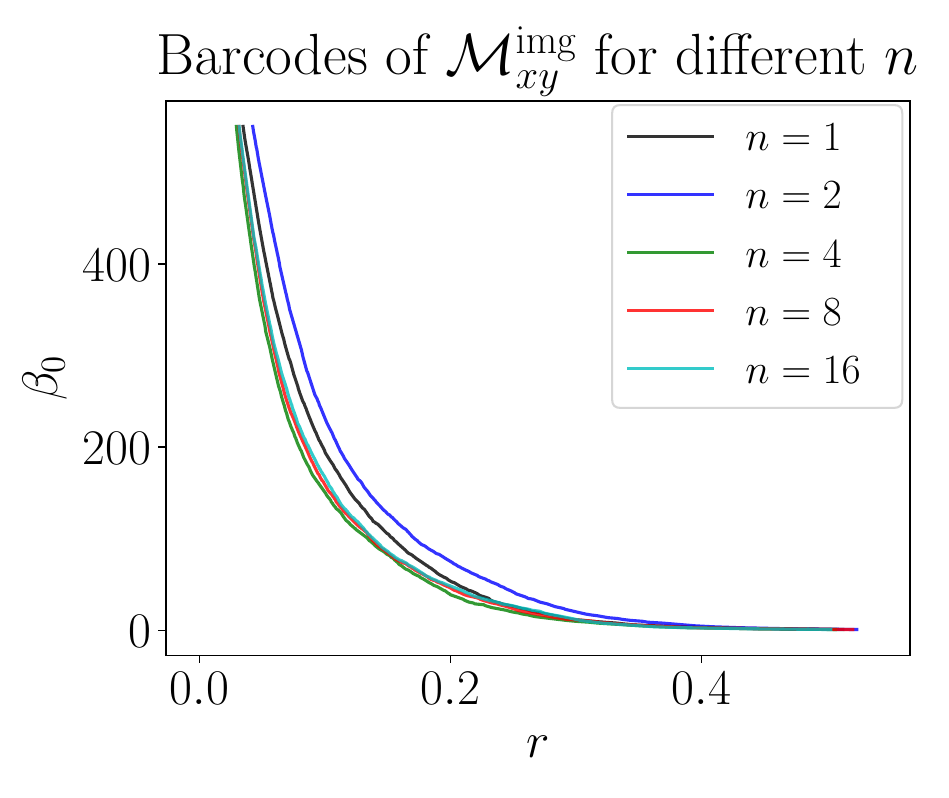}
\put (80,20) {\large{(f)}}
\end{overpic}
\caption{The number of connected components $\beta_0$ of $\op{G}_r(X)$ for different datasets $X$ at different $r$. Pairwise comparison of the persistent barcodes for $\mathcal{M}_{xt,yt}^{\mathrm{res}}$ and  $\mathcal{M}_{xt,yt}^{\mathrm{img}}$ (a), for  $\mathcal{M}_{xy}^{\mathrm{res}}$ and  $\mathcal{M}_{xy}^{\mathrm{img}}$ (b), for $\mathcal{M}_{xy}^{\mathrm{img}}$ and  $\mathcal{M}_{xt,yt}^{\mathrm{img}}$ (c) and for $\mathcal{M}_{xy}^{\mathrm{res}}$ and  $\mathcal{M}_{xt,yt}^{\mathrm{res}}$ (d). Persistent codes of  $\mathcal{M}_{xy}^{\mathrm{img}}$ and $\mathcal{M}_{xt,yt}^{\mathrm{img}}$ for different $n$, (e) and (f). For the sake of visibility, in (e) and (f), only the endpoints of the bars are displayed.}\label{persistent}
\end{figure}
We performed a persistent homology analysis of the manifold to be learned by using \texttt{GUDHI} \cite{gudhi:PersistentCohomology, gudhi:RipsComplex}. We randomly selected 1400 patches of size $18\times 18$, obtaining a set $X\subset \mathbb{R}^{18^2}$ for which we computed its persistent homology. To be able to compare the persistent barcodes at the same scale, we normalized the patches by the maximal pairwise $L_2$-distance of points in $X$. The persistent homology analysis was performed for all patches extracted from the spatio-temporal slices and from spatial image frames by repeating  the experiment ten times  and averaging the obtained number of connected components for each $r\geq 0$ over the experiments. The corresponding barcode diagrams in Figure \ref{persistent} (a) and (b) clearly show that in the spatio-temporal domain as well as in the spatial domain, the residual manifolds are more complex than the manifolds of the ground truth images, i.e. the connected components merge at larger scales $r$. Figure \ref{persistent} (c) also shows that for the ground truth images, the spatial manifold is more complex than the spatio-temporal manifold which is intuitively clear, as the spatial-temporal slices exhibit the temporal correlation of the sequence. This suggests that a network should achieve the best performance when trained to learn the \textit{ground truth spatio-temporal manifold}. Furthermore, we see that in the case of the spatio-temporal domain, the topological complexity tends to be independent of the number of subjects whose patches are extracted to perform the analysis, see Figure \ref{persistent} (c) and (d). In contrast, in the spatial domain, a higher  number of subjects used to extract the patches slightly reduces the topological complexity of the data. Therefore, we conclude that a small number of $2D$ image sequences may already  contain a good representation of all possible two-dimensional spatio-temporal slices  and thus, the number of $2D$ image sequences needed to successfully train a network in the spatio-temporal domain should be lower than for training the network in the spatial domain.
\subsection{Network Architecture}

\begin{figure}[h]
\centering
\begin{overpic}[width=\linewidth,tics=10]{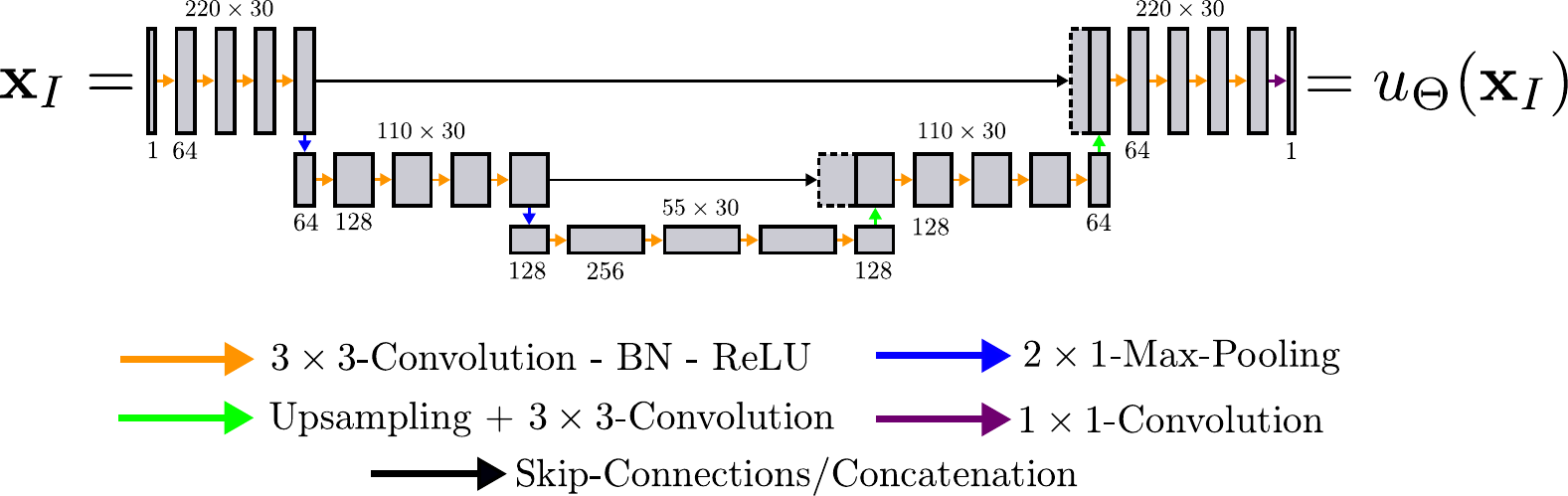}
\end{overpic}
\caption{The U-net with three encoding stages and four convolutional layers per stage, no residual connection and batch-normalization (BN). In the case we train on the spatial domain, max-pooling is performed in both spatial dimensions, whereas in our proposed approach max-pooling is solely performed along the spatial dimension without contracting the data along the temporal dimension. } \label{utheta}
\end{figure}

\begin{figure}[h]
\centering
\begin{overpic}[width=\linewidth,tics=10]{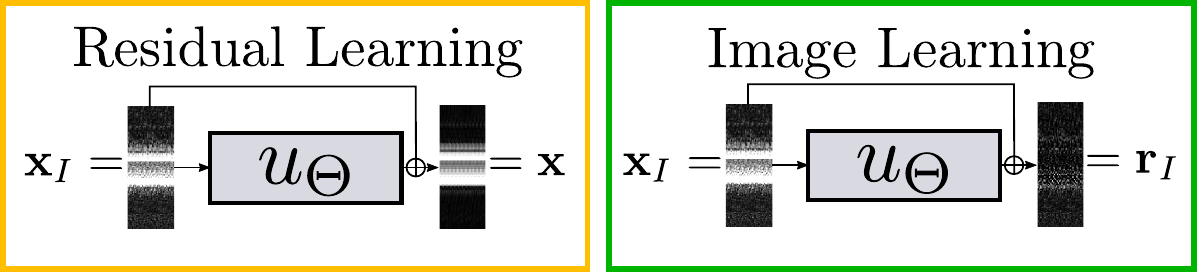}
\end{overpic}
\caption{Residual and Image Learning: For a NN $\tilde{u}_{\Theta}$ with residual connection, learning the residuals is achieved by using the ground truth images $\mathbf{x}$ as labels (left). Learning the ground truth images $\mathbf{x}$ is achieved by using the residuals $\mathbf{r}_I$ as labels (right).}\label{res_connections}
\end{figure}

In the following, we always refer to $\Theta$ as the set of trainable parameters of a network and denote a U-net by $u_{\Theta}$. Figure \ref{utheta} shows the single components of a U-net without residual connection, similar as originally proposed in \cite{ronneberger2015u}. The network consists of five stages, where each stage is a block of four convolutional layers with $2D$ filters of shape $3 \times 3$, followed by batch-normalization \cite{ioffe2015batch} and a component-wise ReLU as activation function. The stages are intercepted by $2\times 1$-max-pooling layers in the encoding phase and by bilinear interpolation layers followed by $3\times 3$ convolutional layers with no activation function in the decoding phase. The initial number of feature maps extracted from the first convolutional layer is set to 64 and is doubled in each block in the encoding phase. The network's output is given by a $1\times1$-convolutional layer which corresponds to a linear combination of the last extracted feature maps. The replacement of the original $2\times 2$-max-pooling by a contraction solely along the spatial dimension empirically turned out to deliver superior results. The black arrows in Figure \ref{utheta} denote concatenations between the last  and the first layer of the corresponding encoding and decoding phases.
\\
Recall from Figure \ref{persistent} in Section \ref{pers_hom_sec} that the manifolds of the ground truth images have a lower topological complexity compared to the manifolds of their corresponding residuals. Therefore, according to \cite{han2016deep} and \cite{bae2017beyond}, one should train the network to learn the features of the artefact-free images. Note that, if the U-net employs a residual connection as in \cite{jin2017deep}, the output is of the form $\tilde{u}_{\Theta}(\mathbf{x}_I) = \mathbf{x}_I + u_{\Theta}(\mathbf{x}_I)$. If $\mathbf{x}$ is used as a label, $\tilde{u}_{\Theta}$ is trained to learn the residual up to a change of sign, as $u_{\Theta}$ is the only part of the network containing trainable parameters. Therefore, being consistent with \cite{han2016deep, bae2017beyond, LeeDeep2017}, in order to exploit the simpler topological complexity of the ground truth images and still be able to benefit from the residual connection as in \cite{jin2017deep}, we propose to train a  U-net with residual connection to estimate the image residuals $\mathbf{r}_I$ of the spatio-temporal slices.
More precisely, if by $\tilde{u}_{\Theta}$ we denote a U-net with residual connection which is trained to map $\mathbf{x}_I$ to the ground truth residuals $\mathbf{r}_I$, and  $\mathbf{r}_{\mathrm{cnn}} = \tilde{u}_{\Theta}(\mathbf{x}_I) = \mathbf{x}_I + u_{\Theta}(\mathbf{x}_I) = \mathbf{x}_I - \mathbf{x}_{\mathrm{cnn}}$, then the estimates of the images are obtained by $\mathbf{x}_I - \mathbf{r}_{\mathrm{cnn}} = \mathbf{x}_I - ( \mathbf{x}_I - \mathbf{x}_{\mathrm{cnn}}) = \mathbf{x}_{\mathrm{cnn}} \approx \mathbf{x}$.\\
Figure \ref{res_connections} shows different approaches for training a U-net to remove undersampling artefacts by training on spatio-temporal slices. Note that, using $\mathbf{x}$ as labels for training a U-net \textit{with}  residual connection and using the residuals $\mathbf{r}_I$ as labels for training a U-net \textit{without} residual connection is equivalent in the sense that the trainable parameters are fitted to learn the residuals $\mathbf{r}_I$. On the other hand, if we want the network to learn the artefact-free images, we can either use the $\mathbf{x}$ as labels and \textit{not employ} a residual connection or use the residuals $\mathbf{r}_I$ as labels and \textit{employ} a residual connection. This holds for training the network on two-dimensional frames as well as on two-dimensional spatio-temporal slices.\\
By $u_{xy}^{\mathrm{res}}$ and $u_{xy}^{\mathrm{img}}$ we denote spatial U-net models when trained to learn the spatial residual manifold $\mathcal{M}_{xy}^{\mathrm{res}}$ and the spatial ground truth image manifold $\mathcal{M}_{xy}^{\mathrm{img}}$, respectively. Analogously, we identify $u_{xt,yt}^{\mathrm{res}}$ and $u_{xt,yt}^{\mathrm{img}}$ as spatio-temporally trained U-nets trained to learn the spatio-temporal manifolds $\mathcal{M}_{xt,yt}^{\mathrm{res}}$ and $\mathcal{M}_{xt,yt}^{\mathrm{img}}$, respectively.

\subsection{Loss Function}
Dependent on what we want the network to learn, we train the network architecture to minimize different loss functions. Let $\mathcal{D}_{xy}^{\mathrm{res}}, \mathcal{D}_{xy}^{\mathrm{img}}$ and $\mathcal{D}_{xt,yt}^{\mathrm{res}}, \mathcal{D}_{xt,yt}^{\mathrm{img}}$ denote the set of available training samples, i.e. the pairs  $(\mathbf{x}_I, \mathbf{r}_I)$ or $(\mathbf{x}_I, \mathbf{x})$, depending on the domain the data is considered in and on which labels are used for training. By $N_{xy}$ and $N_{xt,yt}$ we denote their corresponding cardinality.
Recall that we use the U-net $\tilde{u}_{\Theta}$ to estimate the residual $\mathbf{r}_I = \mathbf{x}_I - \mathbf{x}$ and therefore, the image estimate is given by $\mathbf{x}_{\mathrm{cnn}}= \mathbf{x}_I - \tilde{u}_{\Theta}(\mathbf{x}_I)$. Therefore, in order to define the loss function for a network with residual connection to learn the ground truth images, we use the residuals as labels and vice versa.
The models $u_{xy}^{\mathrm{res}}$ and $u_{xy}^{\mathrm{img}}$ are trained by minimizing the $L_2$-errors between the predicted $2D$ frames and their corresponding labels which are given by
\begin{equation}\label{loss_xy}
\begin{aligned}
\mathcal{L}_{xy}^{\mathrm{res}}(\Theta) &=\frac{1}{N_{xy}} \sum_{(\mathbf{x}_I,\mathbf{x}) \in \mathcal{D}_{xy}^{\mathrm{img}}}\| \tilde{u}_{\Theta}(\mathbf{x}_I) - \mathbf{x}\|_2^2,\\
\mathcal{L}_{xy}^{\mathrm{img}}(\Theta) &=\frac{1}{N_{xy}} \sum_{(\mathbf{x}_I,\mathbf{r}_I) \in \mathcal{D}_{xy}^{\mathrm{res}}}\| \tilde{u}_{\Theta}(\mathbf{x}_I) - \mathbf{r}_I\|_2^2,
\end{aligned}
\end{equation}
respectively. 
In the spatio-temporal case, the models $u_{xt,yt}^{\mathrm{res}}$ and $u_{xt,yt}^{\mathrm{img}}$ are analogously trained   by   minimizing the loss functions 
\begin{equation}\label{loss_xt_yt}
\begin{aligned}
\mathcal{L}_{xt,yt}^{\mathrm{res}}(\Theta) &=\frac{1}{N_{xt,yt}} \sum_{(\mathbf{x}_I,\mathbf{x}) \in \mathcal{D}_{xt,yt}^{\mathrm{img}}}\| \tilde{u}_{\Theta}(\mathbf{x}_I) - \mathbf{x}\|_2^2,\\
\mathcal{L}_{xt,yt}^{\mathrm{img}}(\Theta) &=\frac{1}{N_{xt,yt}} \sum_{(\mathbf{x}_I,\mathbf{r}_I) \in \mathcal{D}_{xt,yt}^{\mathrm{res}}}\| \tilde{u}_{\Theta}(\mathbf{x}_I) - \mathbf{r}_I\|_2^2.
\end{aligned}
\end{equation}

\section{In-Vivo Experiments}\label{section_In_Vivo_Experiments}

\subsection{Data acquisition}

In the following experiments we evaluate the proposed approach on $2D$ Golden radial cine MRI images of 19 subjects (15 healthy volunteers $+$ 4 patients) obtained with a bSSFP sequence on a 1.5T MR scanner (Achieva, Philips Healthcare, Best, The Netherlands) during a 10\,s breathhold (TR/TE = 3.0/1.5\,ms, FA 60$^{\circ}$). 
The spatial dimensions are $N_x \times N_y = 320 \times 320$  with an in plane resolution of 2\,mm and 8\,mm slice thickness. The number of cardiac phases which were reconstructed based on ECG signal is $N_t = 30$. Coil sensitivity information was used to combine the image data of each coil after NUFFT-reconstruction. No further normalization was applied to the image data. The reference images used as ground truth images in the data were reconstructed with $kt$-SENSE \cite{Tsao2003} using $N_{\theta} = 3400$ spokes, which already corresponds to an undersampling factor of $\sim 3$ in each cine image. In addition, dynamic images with $N_{\theta} = 1130$ (3.4\,s scan time) were reconstructed using standard gridding (NUFFT), leading to an undersampling factor of $\sim 9$. 
For each of the 15 healthy volunteers and two patients, $N_z = 12$ slices were acquired while for two patients, only $N_z=6$ slices were obtained due to limited breathhold capabilities. Note that, in contrast to the healthy volunteers, the patients data contains images where the heart movement dysfunction can be diagnosed provided that the temporal information is enough accurate.

\subsection{Evaluation Metrics}

The performance of our method was evaluated in terms of peak signal-to-noise ratio (PSNR), structural similarity index (SSIM) \cite{wang2004image} and  Haar-Wavelet based perceptual similarity index measure (HPSI) \cite{reisenhofer2018haar} as similarity measures and normalized root mean squared error (NRMSE) as error-measure. Note that HPSI has been reported to achieve higher correlation with human opinion scores on different benchmark databases than SSIM \cite{reisenhofer2018haar}.
The quantitative measures are reported for the two-dimensional frames as well as for the two-dimensional spatio-temporal slices after the image sequences were cropped to $160 \times 160 \times 30$ in order to compute the statistics over the regions of interest of the images.

\subsection{Training Set-Up}
Due to our relatively small dataset, all the following experiments were performed in a four-fold cross-validation setting. We split our dataset in portions of 12/3/4 subjects for training/validation/test data, where for one of these configurations, the test data corresponds to the image data coming from patients with heart movement dysfunction.
Obviously, the resulting number of training samples in the spatio-temporal domain is much higher than in the spatial case and therefore, for a fair comparison of the methods, we train the networks by keeping the number of backpropagations fixed. Dependent on the perspective on the dataset, this results in a different number of epochs the networks are trained for. For data-balance reasons, we crop the image sequences using a cut-off of 50 pixels in $x$- and $y$ direction. Therefore, the spatial dimensions per frame reduce to $220\times 220$. Due to the relatively small number of temporal frames and the large receptive field of the U-net, we also conducted experiments evaluating the performance of the networks trained on spatio-temporal slices by mirroring the boundaries. However, as we did not experience any increase or decrease of performance in explicitly handling the boundary conditions, we conducted all experiments on spatio-temporal slices of shape $220 \times 30$. The convolutional layers use zero-padding in order to maintain the spatial shape of the samples constant over each stage. Given a U-net as displayed in Figure \ref{utheta}, we are able to use a mini-batch size of 44 when training in the spatio-temporal domain. Thus, we set the mini-batch size in the spatial training case to 6 in order to have a constant number of pixels  which the networks are fed with per forward pass, i.e. $44 \cdot 220 \cdot 30 =  290\,400= 6 \cdot 220 \cdot 220 $. The networks are trained for $5\cdot 10^4$ backpropagations by stochastic gradient descent (SGD) using a learning rate which was gradually decreased from $10^{-5}$ to $10^{-7}$ and from $10^{-6}$ to $10^{-8}$ for the training in the spatio-temporal domain and in the spatial domain, respectively. The learning rates were chosen in a prior parameter study on the validation set.

\subsection{Residual Vs. Image Learning}

\begin{figure}[!h]		
\centering
\begin{minipage}{\linewidth}
\begin{overpic}
[width=\linewidth,tics=10]{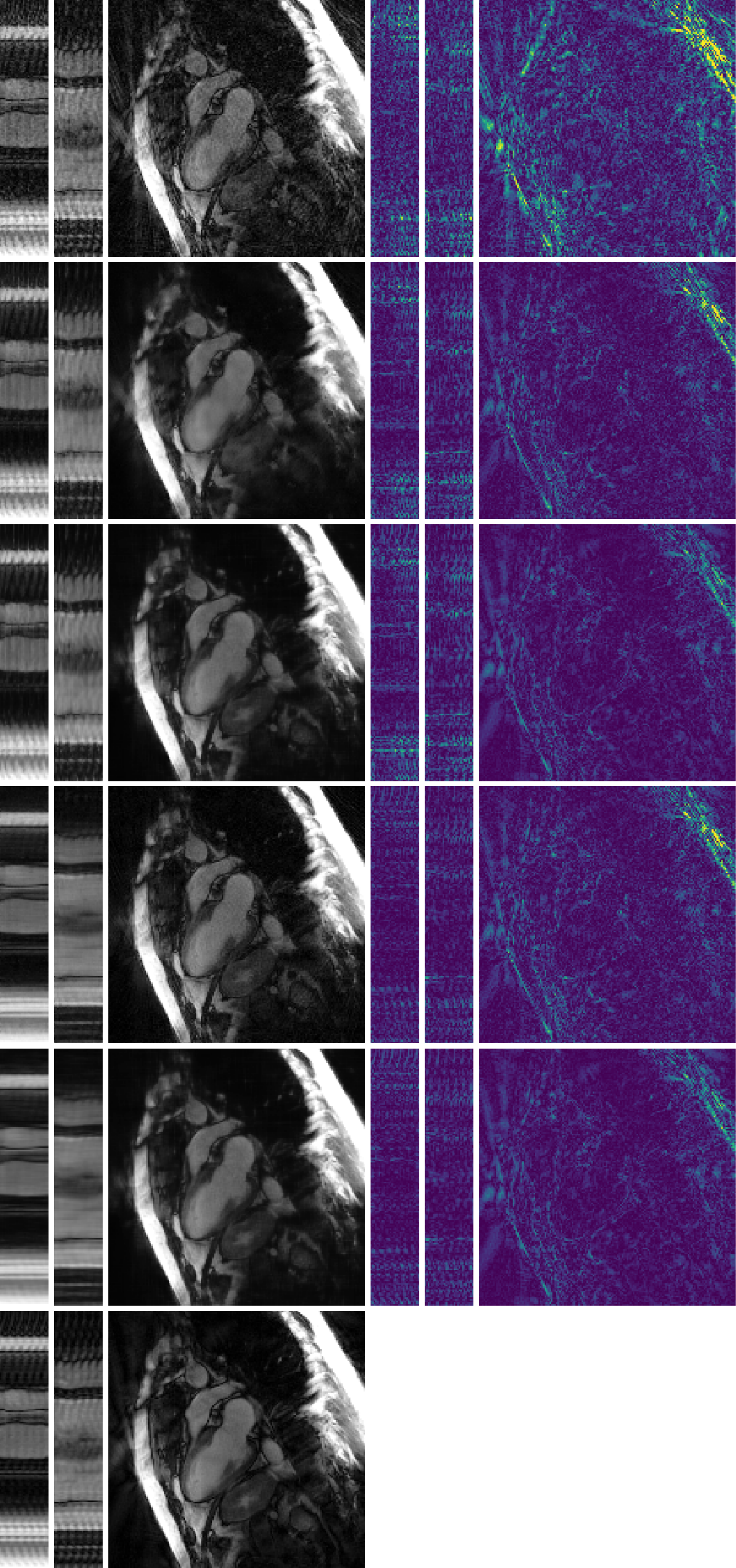}
\put (19.5,85) {\Large\textcolor{white}{(a)}}
\put (19.5,68) {\Large\textcolor{white}{(b)}}
\put (19.5,51.5) {\Large\textcolor{white}{(c)}}
\put (19.5,35) {\Large\textcolor{white}{(d)}}
\put (19.5,18) {\Large\textcolor{white}{(e)}}
\put (19.5,1) {\Large\textcolor{white}{(f)}}
\end{overpic}		  
\end{minipage}
\caption{Comparison of different training approaches for U-nets with residual connection. NUFFT reconstruction with $N_{\theta}=1130$ radial lines (a), spatially trained U-nets $u_{xy}^{\mathrm{res}}$ (b) and $u_{xy}^{\mathrm{img}}$ (c), proposed spatio-temporal approaches $u_{xt,yt}^{\mathrm{res}}$ (d) and $u_{xt,yt}^{\mathrm{img}}$ (e), ground truth (f). The point-wise error images are magnified by a factor of $\times 3$. All images are displayed on the same scale.}\label{res_connections_results}
\end{figure}
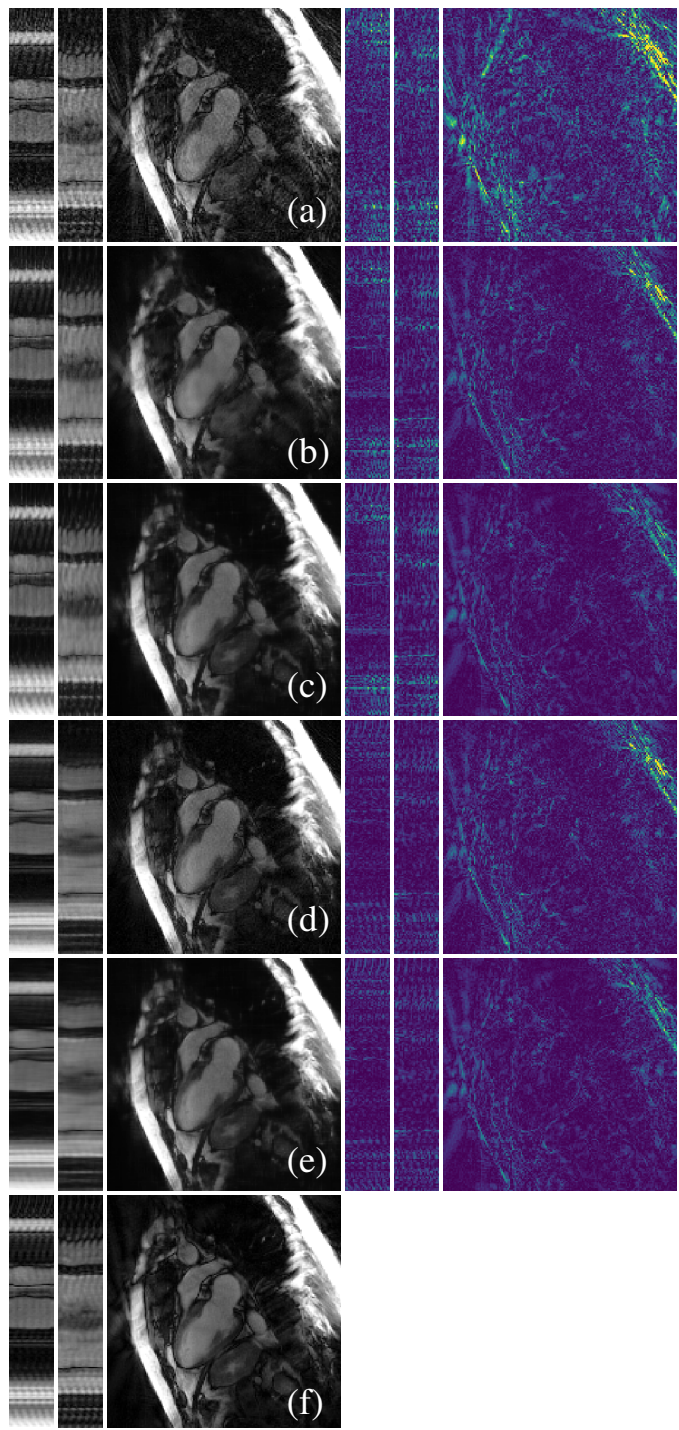

Here we compare the performance of the spatial U-net models $u_{xy}^{\mathrm{res}}$ and $ u_{xy}^{\mathrm{img}}$ and our spatio-temporal approaches $u_{xt,yt}^{\mathrm{res}}$ and $u_{xy}^{\mathrm{img}}$. The models were trained by minimizing the loss functions defined in (\ref{loss_xy}) and (\ref{loss_xt_yt}), respectively.
Figure \ref{res_connections_results} shows qualitative results for different possibilities of training illustrated in Figure \ref{res_connections}.  We see that in both domains, consistent with the previously shown persistent homology analysis, the networks removed the artefacts at their best when they were trained to learn the artefact-free images.  From Figure \ref{res_connections_results} we also already see the superiority of our approach, see (d) and (e), compared to the  spatially trained U-net which slightly tends to smooth out image details and less accurately removed artefacts in spatio-temporal domain, see (b) and (c).
Table \ref{table_RC_tests} shows the results obtained for the spatial U-nets $u_{xy}^{\mathrm{res}}$ and $u_{xy}^{\mathrm{img}}$ and the spatio-temporal U-nets $u_{xt,yt}^{\mathrm{res}}$ and $u_{xt,yt}^{\mathrm{img}}$ for $n=12$, which confirms the heuristics given in Section \ref{pers_hom_sec}. Note that for the experiment, no data-augmentation was used and therefore, the results differ from the ones reported in Table \ref{DL_comparison_table}.
\begin{table}[!h]

\renewcommand{\arraystretch}{1.3}
\begin{center}
\caption{Performance for the spatial and our spatio-temporal approaches dependent on the used architectures.}\label{table_RC_tests}
  \begin{tabular}{l|SS SS}
   \hline
& {$u_{xy}^{\mathrm{res}}$} & {$u_{xy}^{\mathrm{img}} $}  & {$u_{xt,yt}^{\mathrm{res}}$} & {$u_{xt,yt}^{\mathrm{img}}$}\\
\hline
    &  \multicolumn{4}{c}{\textbf{Statistics on $2D$ Frames}}\\

    \textbf{PSNR}  &	34.1203185002 		&	34.714886415475007		&	37.29093834 	&		37.8329612872	\\
     \textbf{SSIM} & 	0.876120508809		&	0.89656138555074993		&	0.927952078453	&		0.93486488827775		\\	
     \textbf{HPSI} & 	0.967856581185		&	0.97920448272100003		&	0.992121406256	&		0.99446704600725		\\	
    \textbf{NRMSE} & 	0.151251760246		&	0.148551926798		& 	0.105501035062	&		0.10517705115525		\\	
   \hline
 &  \multicolumn{4}{c}{\textbf{Statistics on $2D$ Spatio-Temporal Slices}}\\

    \textbf{PSNR}  &	26.4356980973		&	26.4195510687		&	29.4221845321 	&		29.948835674774998	\\
     \textbf{SSIM} & 	0.735186518032		&	0.735130827103		&	0.795267895217	&		0.8040245214185		\\	
     \textbf{HPSI} & 	0.982959604030		&	0.984821113323		&	0.991787699579	&		0.99406181221875		\\	
    \textbf{NRMSE} & 	0.21201634407		&	0.2092147103015		& 	0.160282074985	&		0.1592010417182		\\	

    \hline
  \end{tabular}
\end{center}

\end{table}
As a result, we conclude that for the task of removing undersampling artefacts or image denoising, the relation between the topological complexity of the residuals and the fully-sampled image reconstructions can be used to determine which labels to train the network on as well as how to design the network architecture.
Since the radial acquisition is designed to be incoherent along the temporal dimension, in all our following experiments we use the U-net architecture as shown in Figure \ref{utheta} where we make use of the residuals as labels and employ a residual connection as shown in Figure \ref{res_connections} for the case of image learning. In the next Subsection, we also see how learning the manifold $\mathcal{M}_{xt,yt}^{\mathrm{img}}$ can reduce the training time as convergence of the training and validation errors is achieved faster.

\subsection{Training with Limited Amount of Data}
Here we demonstrate the performance of our proposed approach when we restrict the number of available training samples. For this purpose, we trained the same network on different datasets where we fixed a different number of subjects $n$ whose images we included in the training dataset. We show that with our proposed approach we are able to obtain comparable results even with a small number of subjects.

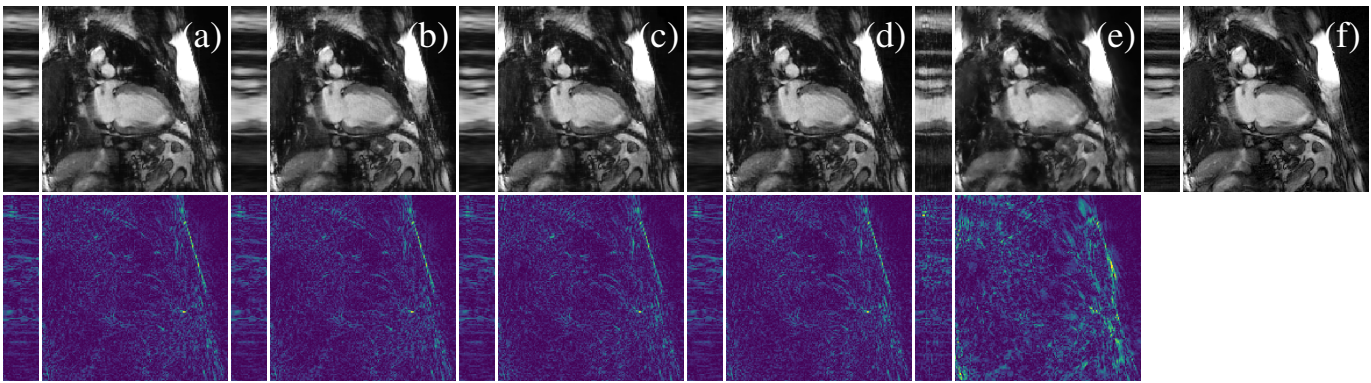
\begin{figure*}[t]				
\begin{overpic}
[width=\linewidth,tics=100]{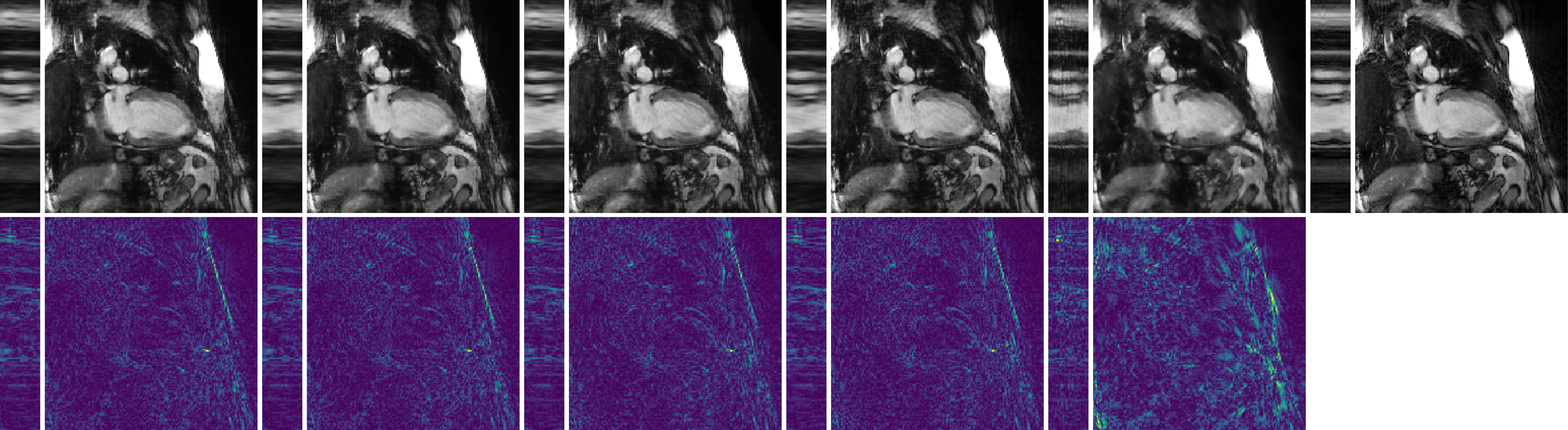}
\put (13,24.5) {\Large\textcolor{white}{(a)}}
\put (29.5,24.5) {\Large\textcolor{white}{(b)}}
\put (46.5,24.5) {\Large\textcolor{white}{(c)}}
\put (63,24.5) {\Large\textcolor{white}{(d)}}
\put (80,24.5) {\Large\textcolor{white}{(e)}}
\put (96.75,24.5) {\Large\textcolor{white}{(f)}}
\end{overpic}	
\caption{Results on the test set for $N_{\theta} = 1130$ radial lines when the number of subjects whose spatio-temporal slices are extracted was varied. Note that
no data-augmentation was used. Proposed method for $n = 1$ (a), $n =2$ (b), $n = 8$ (c), $n = 12$ (d), the spatial U-net for $n=12$ (e) and the $kt$-SENSE reconstruction with 3400 radial lines (f). The point-wise error images are magnified by a factor of $\times 3$. All images are displayed on the same scale.}\label{n_patients_variation}
\end{figure*}
Note how in the spatial training scenario, the given training data is naturally constrained by the fact that for a fixed slice, different time frames of the ground truth images exhibit a high similarity. Therefore, regardless of the fact that in the spatial domain the ground truth image manifold has a lower complexity than the residual manifold, a network which is trained to learn the ground truth images should be expected to suffer from the limited variability of the data. In contrast, due to the temporal incoherence of the undersampling pattern, this issue should be overcome when learning the residuals.
In the spatio-temporal domain, the availability of the data is not an issue as we have $n\,N_z (N_x + N_y) \gg n\, N_z\,N_t$ samples. Therefore, one would expect the performance of the network to be to some extent independent of the number of subjects $n$ the samples are extracted from. Also, according to the performed persistent homology analysis, the training of the network should be facilitated when trained to learn the manifold of the ground truth images.

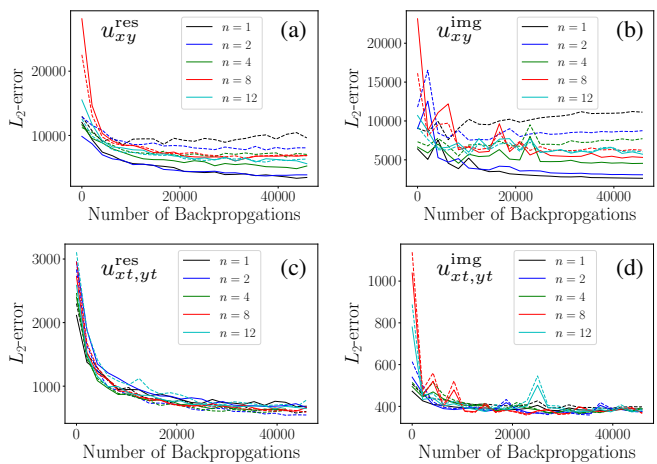
\begin{figure}[!h]			
\centering
\begin{overpic}[width=0.49\linewidth,tics=10]{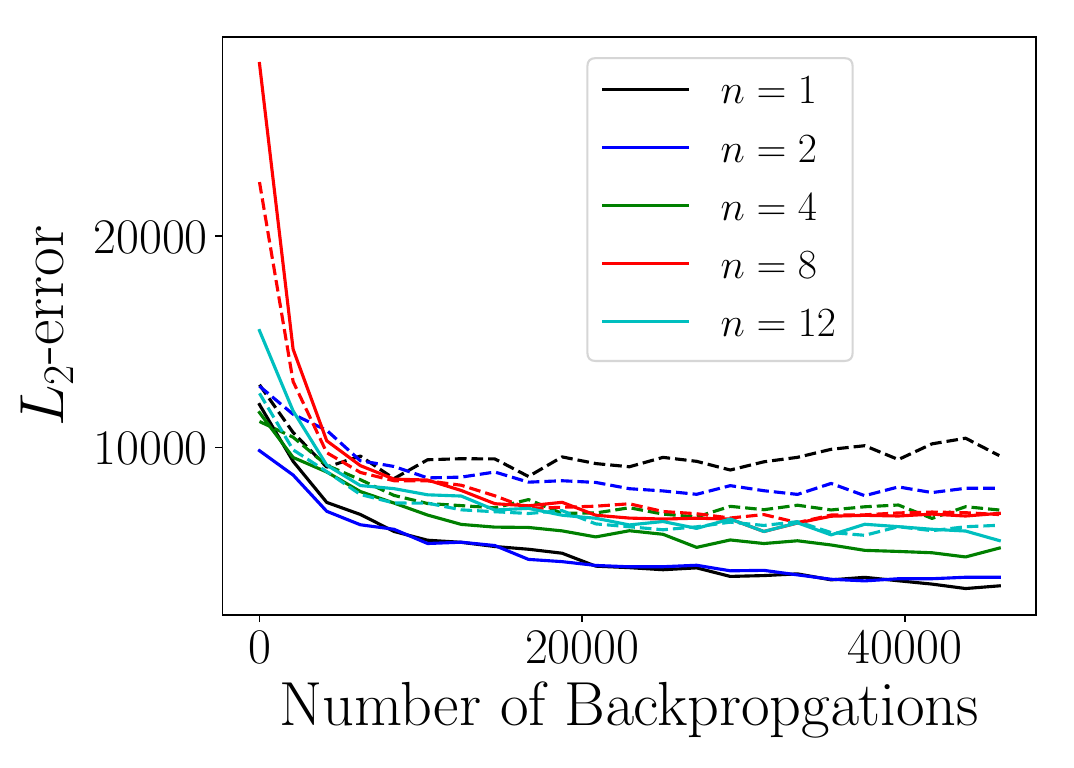}\hspace{0.2cm}
\put (85,60) {\small\textcolor{black}{(a)}}
\put (30,60) {\small\textcolor{black}{$u_{xy}^{\mathrm{res}}$}}
\end{overpic}
\begin{overpic}[width=0.49\linewidth,tics=10]{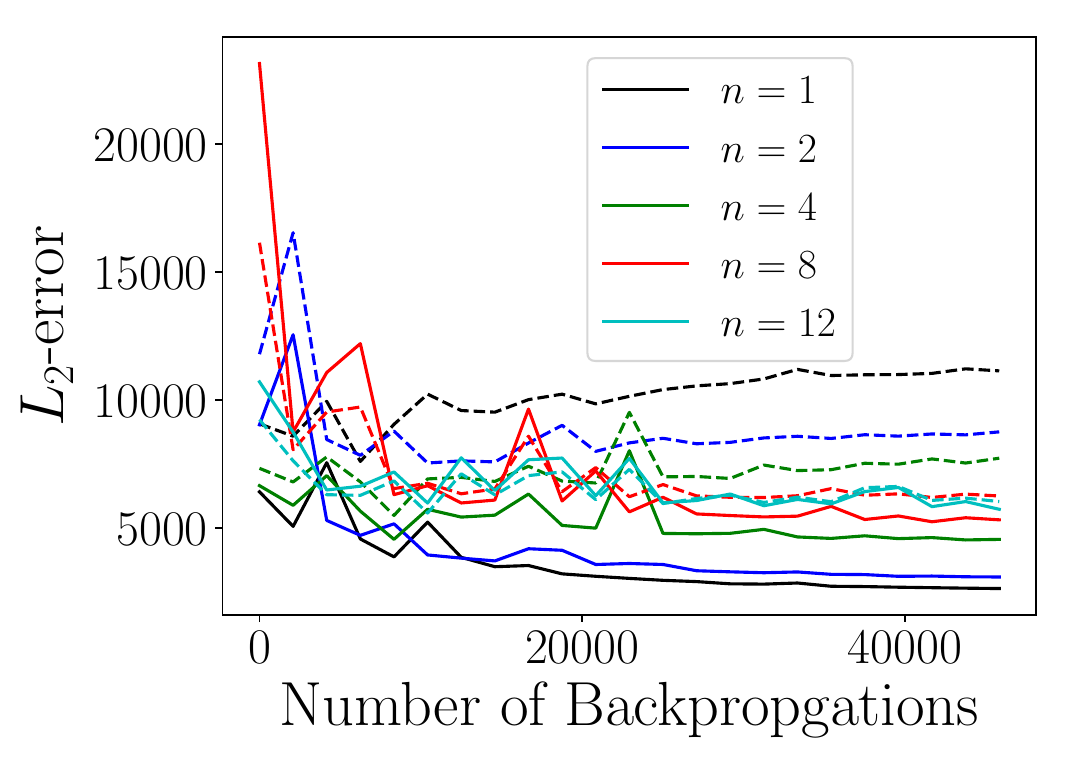}\hspace{0.2cm}
\put (85,60) {\small\textcolor{black}{(b)}}
\put (30,60) {\small\textcolor{black}{$u_{xy}^{\mathrm{img}}$}}
\end{overpic}
\begin{overpic}[width=0.49\linewidth,tics=10]{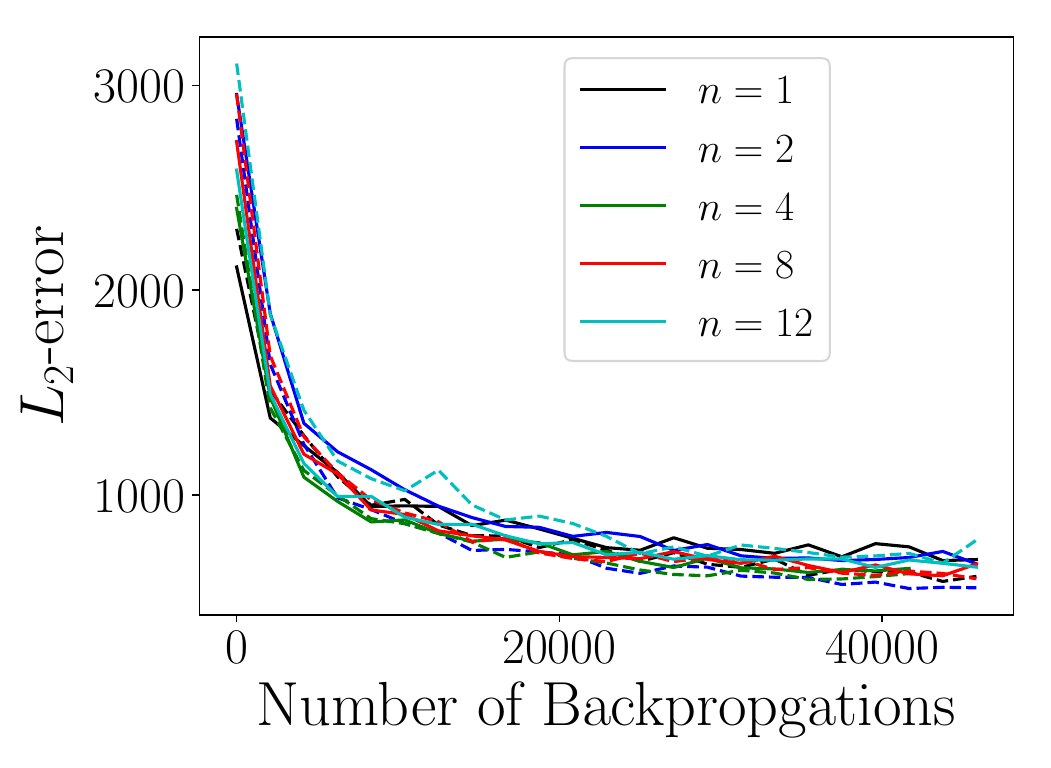}\hspace{0.2cm}
\put (85,60) {\small\textcolor{black}{(c)}}
\put (30,60) {\small\textcolor{black}{$u_{xt,yt}^{\mathrm{res}}$}}
\end{overpic}
\begin{overpic}[width=0.49\linewidth,tics=10]{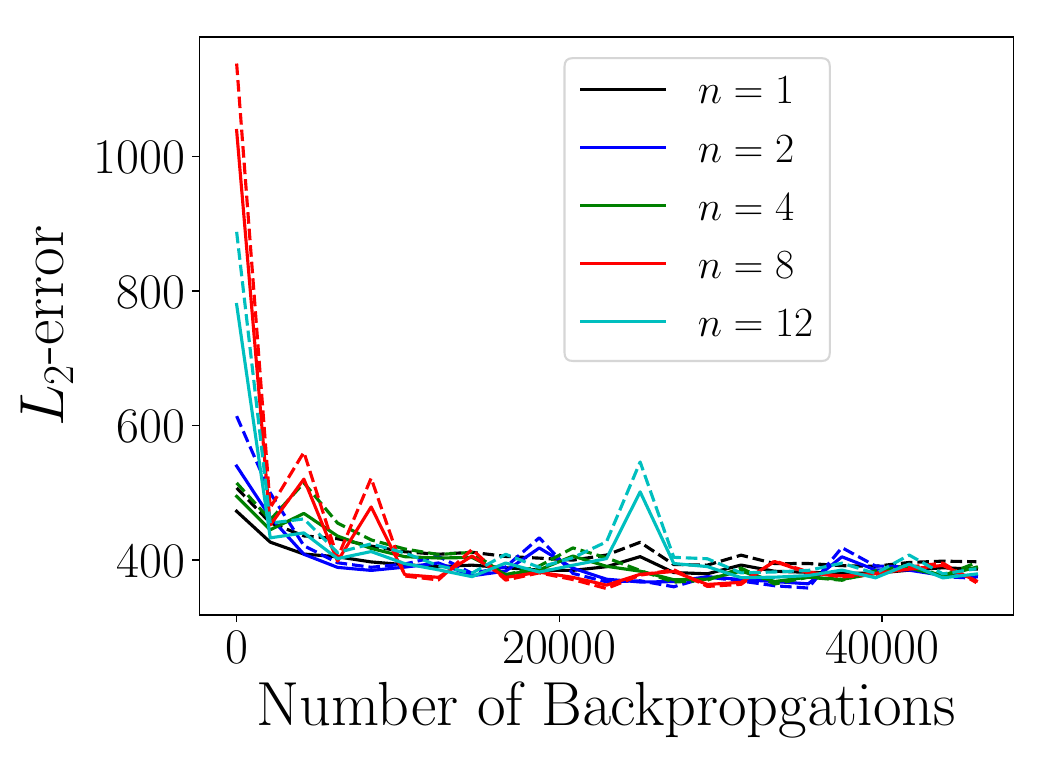}\hspace{0.2cm}
\put (85,60) {\small\textcolor{black}{(d)}}
\put (30,60) {\small\textcolor{black}{$u_{xt,yt}^{\mathrm{img}}$}}
\end{overpic}
\caption{Loss behaviour during training with $N_{\theta}=1130$ for different number of volunteers $n$ contained in the dataset. Training loss (solid) and validation loss (dashed) for the spatial and spatio-temporal U-nets. Spatial residual learning (a), spatial image learning (b), spatio-temporal residual learning (c), spatio-temporal image learning (d). Note that the scales differ due to the different losses and the different domains in which the networks are trained.}
\label{t_v_losses}
\end{figure}
Figure \ref{t_v_losses} shows the behaviour of the loss decay for the spatial approach ((a) and (b)), the spatio-temporal training approach ((c) and (d)), and in both cases, for the situation where the residuals are learned ((a) and (c)) and where the ground truth images are learned ((b) and (d)). We see that for the spatial U-net, for the residual learning and the image learning, increasing the number of subjects $n$ leads to a decrease of the gap between training and validation error. Further, we see that the gaps are larger in the case where the ground truth images are learned which can be related to the low variability of the dataset. In both cases, for $n=12$ the gap is small enough to assume that the networks have been properly trained and generalize well. 
For $n=1$ and for $n=1,2,4$, the spatially trained U-nets  $u_{xy}^{\mathrm{res}}$ and  $u_{xy}^{\mathrm{img}}$ poorly generalize in both training scenarios, as the networks almost immediately start to overfit the data, see (a) and (b). Spatial training of the networks without data-augmentation is possible for $n=2,4,8,13$ for the residual learning and for $n=8,13$ for the image learning. However, our method outperforms the spatially trained U-net as it better maintains diagnostic details in spatial and spatio-temporal domain, see Figure \ref{n_patients_variation} for the case $n=12$. For the spatio-temporal approaches, the gaps between training and validation error are smaller compared to the ones for the spatial approaches. This holds for the residual learning as well as the image learning scenario. Further, when the network is trained to learn the ground truth images, the errors converge faster than in the residual training approach, compare Figure \ref{t_v_losses} (c) and (d). Also, the convergence rate is highly independent on the number of subjects $n$.
From these experiments, we first conclude that our proposed method is well suited for training a network on a limited number of subjects. Second, forcing the network to learn the manifold given by the ground truth images  $\mathcal{M}_{xt,yt}^{\mathrm{img}}$  facilitates the training, which leads to a faster convergence of the errors and therefore to lower training times.
\begin{table}

\renewcommand{\arraystretch}{1.3}
\begin{center}
\caption{Results on the test when the number of subjects whose images were included in the training set is varied.}\label{table_n_patients_variation}
  \begin{tabular}{l|SSSSS}
   \hline
& {$n =1 $} & {$n =2 $} &{$n =4 $} & {$n =8 $} & {$n =12 $}\\
\hline
    
    &  \multicolumn{5}{c}{\textbf{Statistics on $2D$ Frames}}\\
    \textbf{PSNR} &	 37.2454621043		&	37.7852403246		&	37.6588933875 	&	37.844872852449996		&	37.8329612872		\\
     \textbf{SSIM} & 0.931165370465		&	0.9336472447494999	&	0.933997787489	&	0.93447901849475		&	0.93486488827775	\\	
     \textbf{HPSI} & 0.9935205122040001	&	0.9942556093020001	&	0.99424253870475	&	0.99444056531425	&	0.99446704600725	\\	
    \textbf{NRMSE} & 0.109304008118		&	0.10597785489422501	&	0.10723638026925	&	0.10513810543675	&	0.10517705115525	\\	
   \hline
   
 &  \multicolumn{5}{c}{\textbf{Statistics on $2D$ Spatio-Temporal Slices}}\\
\textbf{PSNR} &	 29.583765368075	&	29.90118016455		&	29.773929414125003	&	29.951796792475		&	29.948835674774998	\\   
\textbf{SSIM} &	 0.79344640854125	&	0.80062882150125	&	0.80153708372525	&	0.80334640239775	&	0.8040245214185	\\   
  \textbf{HPSI} & 0.99292217079025	&	0.9938650912635		&	0.99380728677975	&	0.99403761366825	&	0.99406181221875	\\ 
\textbf{NRMSE} & 0.16040829522975	&	0.15977596646750003	&	0.1621793371135		&	0.16078637836		&	0.1592010417182	\\   
    \hline
  \end{tabular}
\end{center}

\end{table}
Figure \ref{n_patients_variation} shows a slice  of the output of an image in the test set which was obtained with our proposed method. For all $n$, the artefacts have been successfully removed. We also see that even for $n=1$, the dataset is already rich enough in order to allow for a good depiction of cardiac contraction and expansion during the heart cycle. Table \ref{table_n_patients_variation} shows the achieved average of the quantitative measures. Even if in terms of quantitative measures the network performs better the larger the training data, the differences are marginal and hardly perceivable by the human eye, see Figure \ref{n_patients_variation}. Therefore, we conclude that since the data has a particularly simple structure, little data is already sufficient for a successful training.

\subsection{Rotation Equivariance}
\begin{figure}[H]
\centering
 \begin{overpic}[height=0.4\linewidth,tics=10]{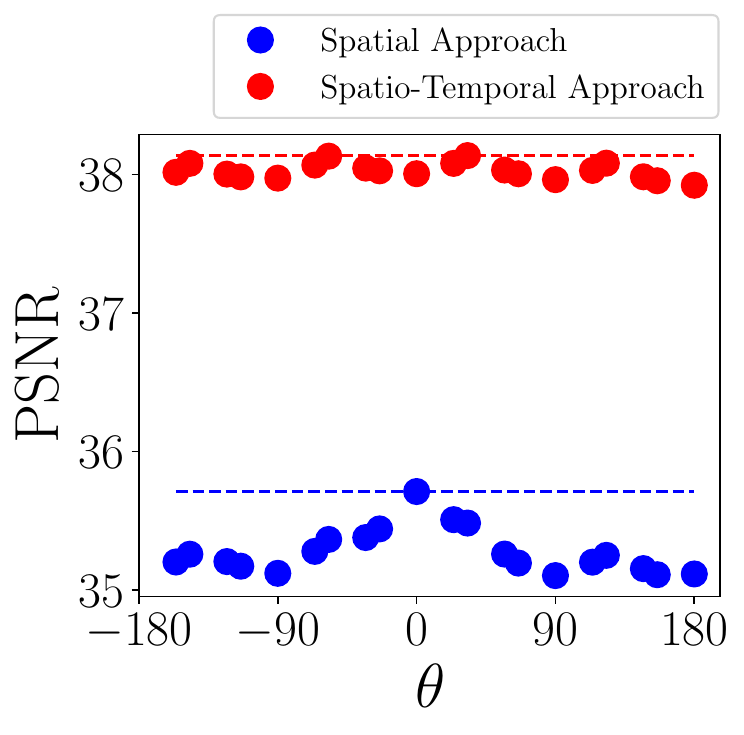}

\end{overpic}
\hspace{0.1cm} \begin{overpic}[height=0.4\linewidth,tics=10]{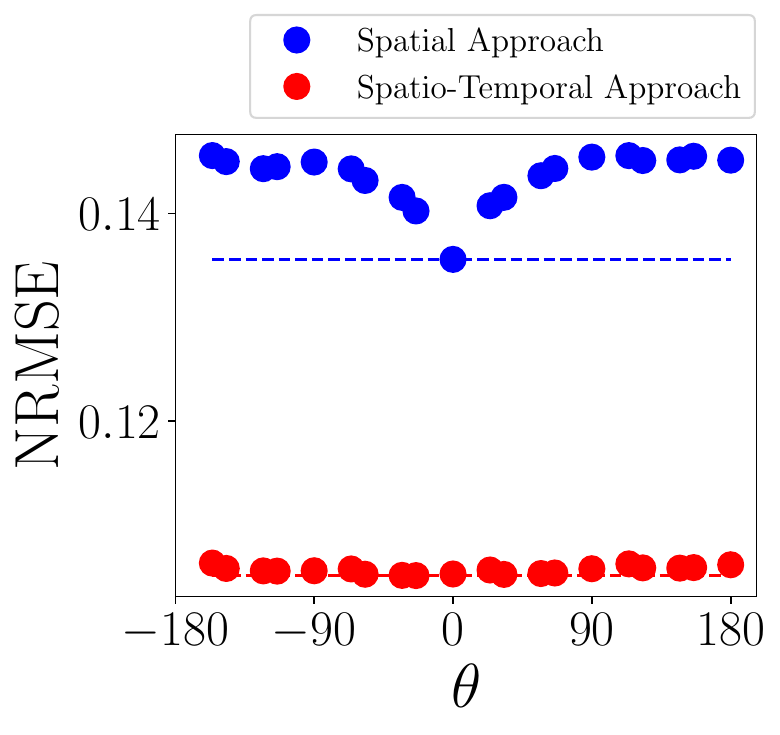}

\end{overpic}\\ 
\vspace{0.2cm}
 \begin{overpic}[height=0.4\linewidth,tics=10]{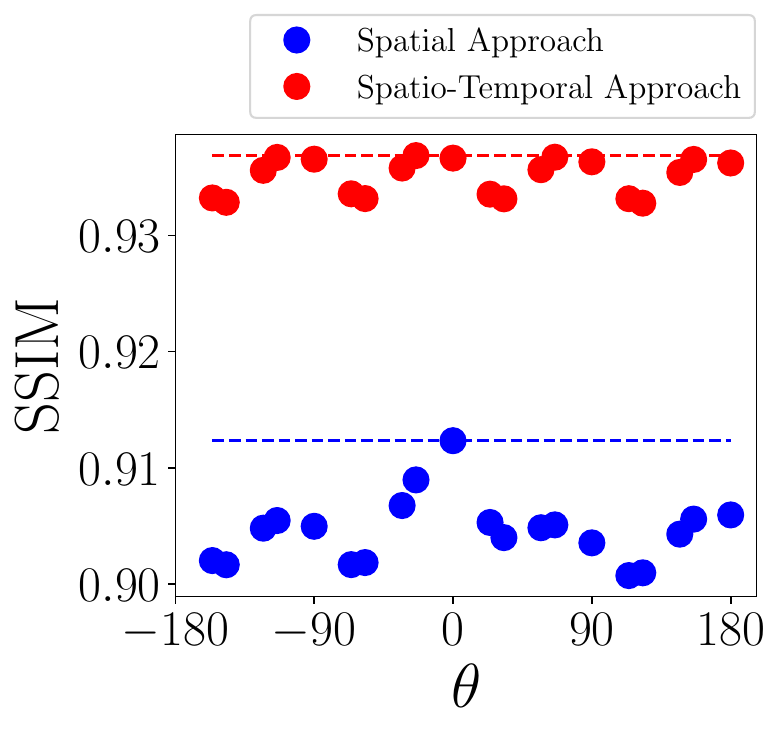}

\end{overpic}
 \begin{overpic}[height=0.4\linewidth,tics=10]{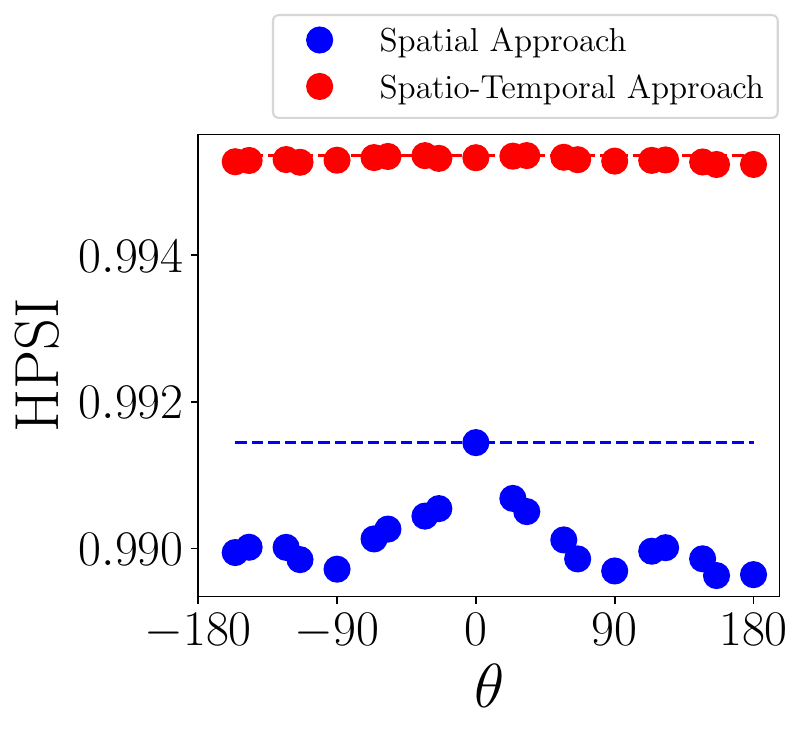}

\end{overpic}
\caption{Performance of the networks when tested on rotated copies of the images contained in the training set. While the network trained in the spatio-temporal domain is robust with respect to rotation, the network trained on images in the spatial domain loses generalization power when tested on rotated copies of the images it was trained on. The dashed lines correspond to the corresponding measure achieved on the training set.}\label{rot_figures}
\end{figure}

CNNs are well known to be able to achieve properties as translation-invariance and -equivariance \cite{goodfellow2016deep}. However, they are not naturally invariant or equivariant with respect to rotation and one of the still most used methods to achieve these properties is to appropriately augment the dataset, \cite{krizhevsky2012imagenet, litjens2017survey}. In contrast, other approaches \cite{worrall2017harmonic}, \cite{marcos2017rotation}, \cite{cohen2016group} explicitly incorporate invariant/equivariant convolutional operations in the networks which comes at the cost of a more complex network design.
As a rotation in image space, i.e. due to a rotation of the field of view in order to adapt the scan to the geometry of the patient's heart, might easily be encountered, we are interested in achieving rotation-equivariance, i.e. $f_{\Theta}(\psi(\mathbf{x}_I)) = \psi(f_{\Theta}(\mathbf{x}_I))$ for an already trained network $f_{\Theta}$ and rotation $\psi$ in the $xy$-plane. For the following experiment, we generated new different test sets $\mathcal{D}_{xy}^{\psi_{\theta}}$ and $\mathcal{D}_{xt,yt}^{\psi_{\theta}}$ by applying rotations $\psi_{\theta}$ with rotation-angle $\theta$ and tested the networks which were previously trained on the non-rotated images on the different test sets. By doing so, we were able to isolate and measure the direct effect of the sole rotation in image space on the performance of the network.\\ 
We rotated the measured data in $k$-space and reconstructed the training set for different angles $\theta$. Note that the process is time consuming since the images were reconstructed with $kt$-SENSE. Therefore, we only reconstructed rotated images for $\theta = \pm 66^{\circ},\pm 33^{\circ}$ and for each $\theta$ we further rotated the frames by $\pm90^{\circ}$ and $180^{\circ}$, obtaining an overall number of 19 rotated test sets. 
Figure \ref{rot_figures} compares our approach to the $2D$ spatially trained U-net in terms of quantitative measures calculated over the $2D$ frames of the different test sets with different rotation angles. For $\theta=0$, the measures indicate the average measure achieved on the training set. First, we see again that the spatio-temporal training approach clearly outperforms the spatial training approach in terms of all quantitative measures. Further, while rotating the $2D$ frames yields a noticeable decrease of performance of the network  trained in the spatial domain, the network trained on the spatio-temporal slices performs similarly well on the different rotated test sets and is therefore almost rotation-equivariant. 

\subsection{Experiments with Shallower Networks}

\begin{figure}[!h]
\centering
\begin{overpic}[width=\linewidth,tics=3]{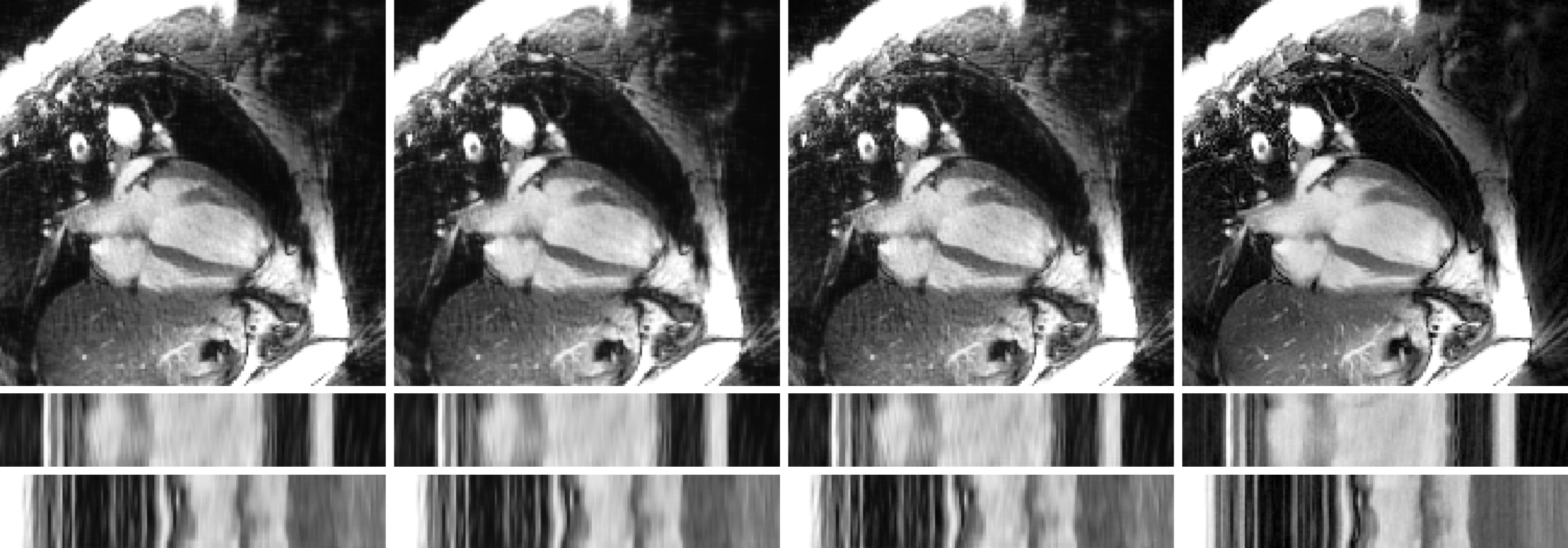}
 \put (17,29) {\Large\textcolor{white}{(a)}}
 \put (42,29) {\Large\textcolor{white}{(b)}}
 \put (67,29) {\Large\textcolor{white}{(c)}}
 \put (92,29) {\Large\textcolor{white}{(d)}}
\end{overpic}
\caption{Results obtained with different CNNs following our proposed aproach $u_{xt,yt}^{\mathrm{img}}$. E1\,C8 (a), E4\,C4 (b) and E5\,C2 (c), $kt$-SENSE reconstruction with $N_{\theta} = 3400$ radial lines (d). Our approach therefore offers the possibility to further reduce the network complexity as well as training times.}\label{different_architectures_results}
\end{figure}
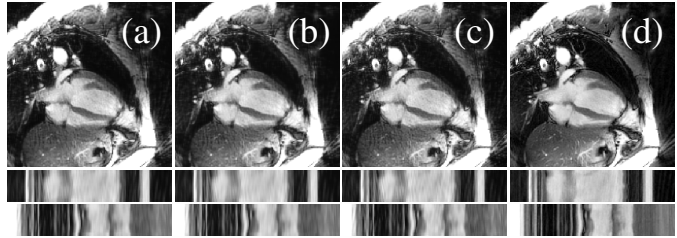

Even if we used the network architecture shown in Figure \ref{utheta} for all experiments, the strength of the method lies in the change of perspective on the data. To demonstrate this, we applied different network architectures following our suggested approach. More precisely, we tested different types of CNNs which can be seen as special cases of the U-net. If by E and C we denote the numbers of encoding stages and convolutional layers per stage of a U-net, E3\,C4 corresponds to the network displayed in Figure \ref{utheta}. E1\,C8, on the other hand, denotes a single-scale fully CNN with eight convolutional layers and no max-pooling.
Figure \ref{different_architectures_results} shows results obtained with different network architectures parametrized by E and C. We see that the networks E1\,C8 and E4\,C4 which differ in terms of number of trainable parameters by approximately a factor of 10, achieve similar performance. This suggests that the number of trainable parameters and consequently, also training times, could further be reduced without significantly losing performance. Figure \ref{different_architectures_results}  shows results obtained by E1\,C8 (a), E4\,C4 (b) and E5\,C2 (d), where the networks were trained for $3\cdot 10^4$ backpropagations. The training of E1\,C8, for example, see Figure \ref{different_architectures_results} (a), amounted to only 40 minutes.

\subsection{Comparison with other Deep Learning-based Methods}
\begin{figure}[!b]				
\begin{minipage}{\linewidth}
\centering
\begin{overpic}[width=\linewidth,tics=3]{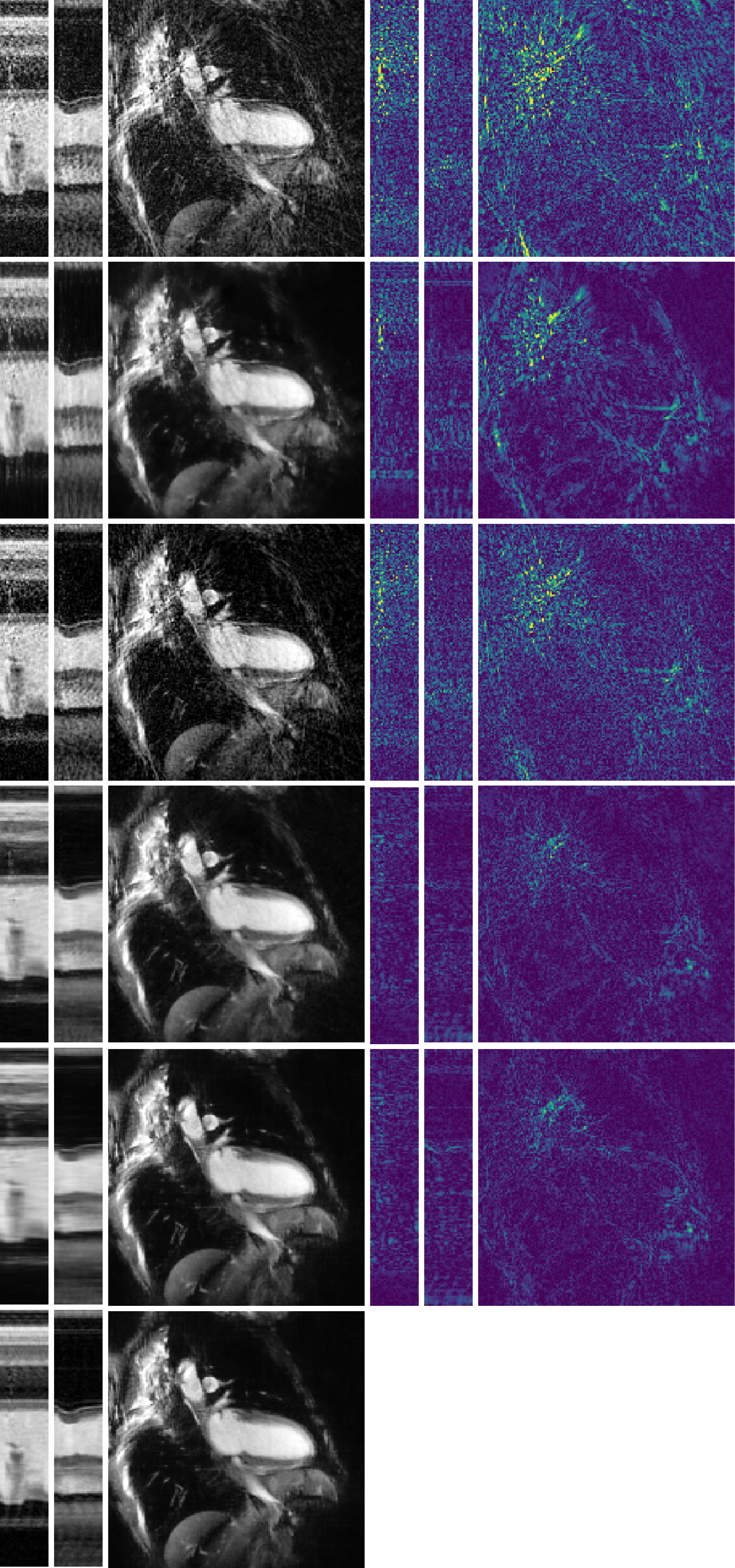}
\put (19.5,85) {\Large\textcolor{white}{(a)}}
\put (19.5,68) {\Large\textcolor{white}{(b)}}
\put (19.5,51.5) {\Large\textcolor{white}{(c)}}
\put (19.5,34.5) {\Large\textcolor{white}{(d)}}
\put (19.5,18) {\Large\textcolor{white}{(e)}}
\put (19.5,1) {\Large\textcolor{white}{(f)}}
 \end{overpic}
\end{minipage}
\caption{Comparison with different Deep Learning-based post-processing methods. NUFFT reconstruction with $N_{\theta} = 1130$ radial lines (a), $u_{xy}^{\mathrm{img}}$ (b), $u_{xy,t}$ (c), $u_{xyt}$ (d), proposed approach $u_{xy}^{\mathrm{img}}$ (e), ground truth $kt$-SENSE reconstruction (f). The point-wise error images are magnified by a factor of $\times 3$. All images are displayed on the same scale.}\label{DL_comparison_fig}
\end{figure}
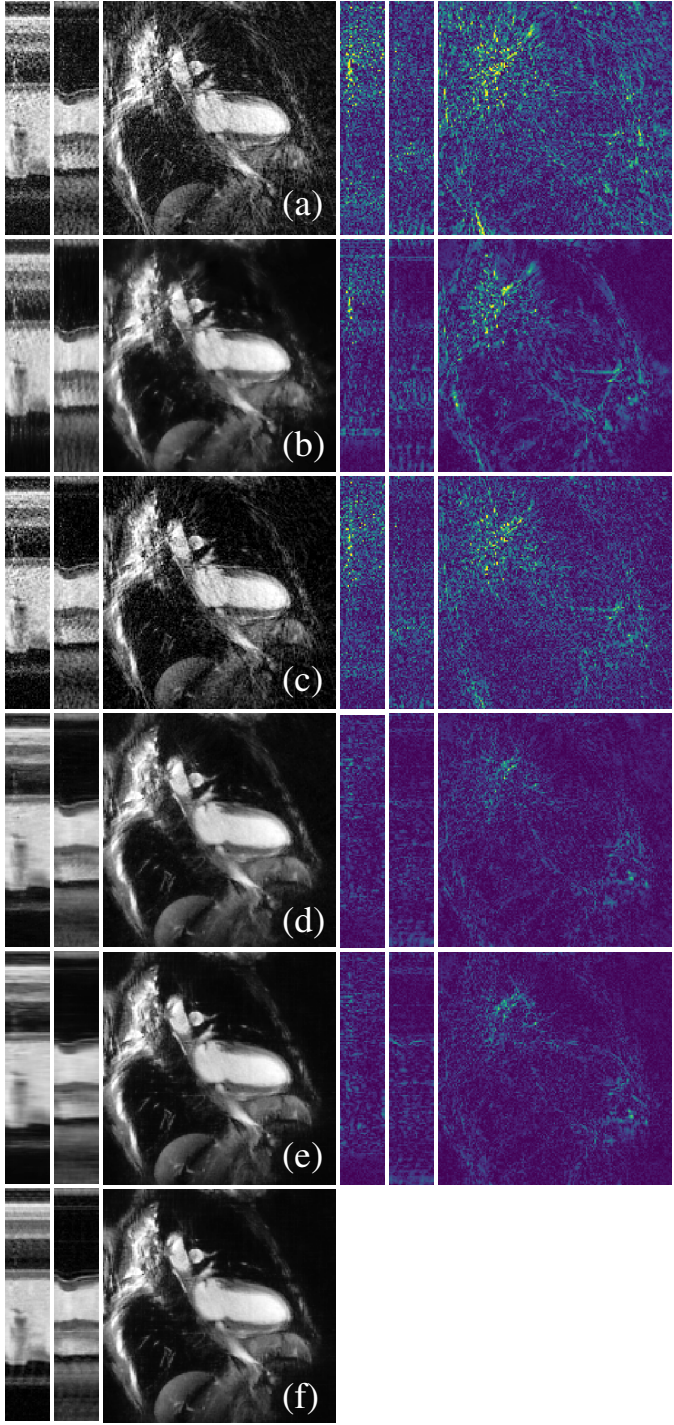

Here we compare our approach to other methods based on post-processing with deep NNs. Since we only have access to a limited dataset, for the following experiments, we made use of data-augmentation by using all our rotated images, flipping, shifting the images along the channel axis and adding random constant values to the whole image sequences. By doing so, we created a potentially infinite training set. Note that we did not include elastic deformations as a data-augmentation technique, as the data-acquisition process is not simulated and elastic deformations might alter the structure of the undersampling artefacts in the input data.
The first method of comparison is the already discussed spatially trained U-net $u_{xy}^{\mathrm{img}}$. It is trained to map frames to frames and corresponds to the method discussed in \cite{jin2017deep} and \cite{han2016deep}. The second method of comparison is a natural extension of the first and corresponds to the $2D$ U-net approach shown in Figure \ref{different_approaches} (b) which we refer to as $u_{xy,t}$. The net is trained to map whole image sequences to whole image sequences by aligning the cardiac phases along the channel's axis and was presented in \cite{sandino2017deep}. Further, we compare our method to the $3D$ U-net approach $u_{xyt}$ presented in \cite{Hauptmann2019}, see Figure \ref{different_approaches} (c).
While for the $2D$ NNs, we cropped the images to $220 \times 220$ and $220 \times 220 \times 30$ in order to let the networks focus on the diagnostic content of the images, for the $3D$ U-net, the images used for training needed to be cropped to $128 \times 128 \times 20$, as the network is computationally more expensive. The shape was the one used in \cite{Hauptmann2019}. In order to obtain image sequences of $320 \times 320 \times 30$, the outputs of the networks were treated as patches and the image sequences were reconstructed from the patches by properly averaging over regions with overlapping patches. In contrast to the models employing $2D$ convolutional layers, which were trained using SGD, the $3D$ U-net $u_{xyt}$ was trained in the same setting as suggested in \cite{Hauptmann2019} using \textit{ADAM} \cite{kingma2014adam}. Figure \ref{DL_comparison_fig} and Table \ref{DL_comparison_table} show and summarize the obtained results with the described networks. For more detailed information about the reassembling of the image sequences from the patches, see Section \ref{reco_time_subsec}.

\begin{table}[!h]

\renewcommand{\arraystretch}{1.3}

 \caption{Comparison of different Deep Learning-based post-processing approaches.}\label{DL_comparison_table}
\centering

\begin{tabular}{l|S S S S }
   \hline
     \textbf{NN Model}  &  {\textbf{$u_{xy}^{\mathrm{img}}$}} & {\textbf{ $u_{xy,t}$ }}    &{\textbf{$u_{xyt}$}} &   {\textbf{$u_{xt,yt}^{\mathrm{img}}$}}   \\
       \hline
       &  \multicolumn{4}{c}{\textbf{Statistics on $2D$ Frames}}\\

    \textbf{PSNR}  & 34.8171640121	 	& 33.52628051115	 	&  37.827453633350004  &   	  37.929795124775	 	\\
     \textbf{SSIM} &   0.910299657663		& 0.8684890288452499 	&  0.9354742965980001   &    0.934907907027	\\
     \textbf{HPSI} & 0.98767046754725007		&  0.9835626116375		&   0.9944677153085    &    0.99449713972625	 \\
    \textbf{NRMSE} & 0.14119983992975		&  0.17238043159549998	&   	0.1051231439705	 &	   0.10420685048975	\\
   \hline
     & \multicolumn{4}{c}{\textbf{Statistics on $2D$ Spatio-Temporal Slices}}\\

    \textbf{PSNR}  & 26.959143560400001	 	&	25.237788508225  	&  29.872686641625   & 	     30.04809129495 	 	\\
     \textbf{SSIM} & 0.73997698041449989	&  0.6932218916850001   &   0.808525650412  & 	    0.8037551523705 	\\
     \textbf{HPSI} & 0.990355946322	& 0.9905011649467501  	&  0.9941683563275001	  	 &  0.994095040354   	 \\
    \textbf{NRMSE} & 0.208035677671	 & 0.289838227941  		&  0.1651662001145 &	    0.15842918636375  	\\
    \hline
  \end{tabular}
\end{table}
\begin{figure}[!h]
\begin{minipage}{\linewidth}
\centering

 \begin{overpic}[width=0.49\linewidth,tics=10]{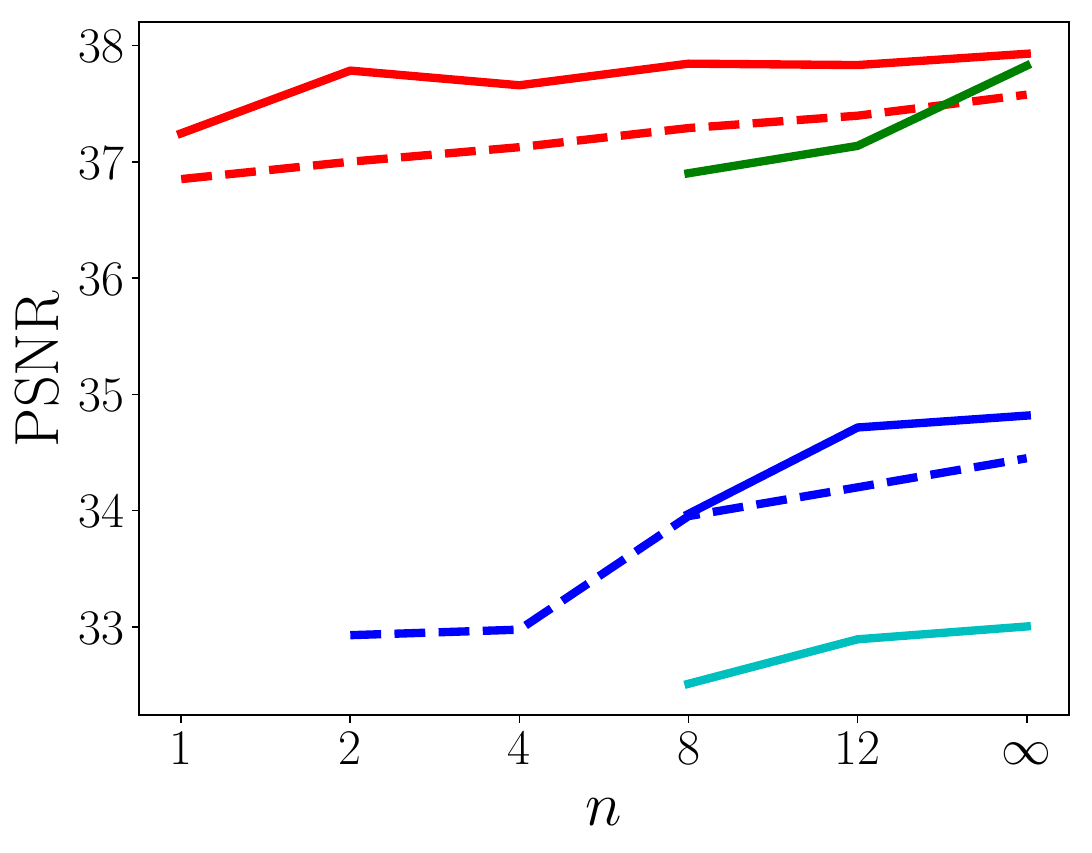}

\end{overpic}
\begin{overpic}[width=0.49\linewidth,tics=10]{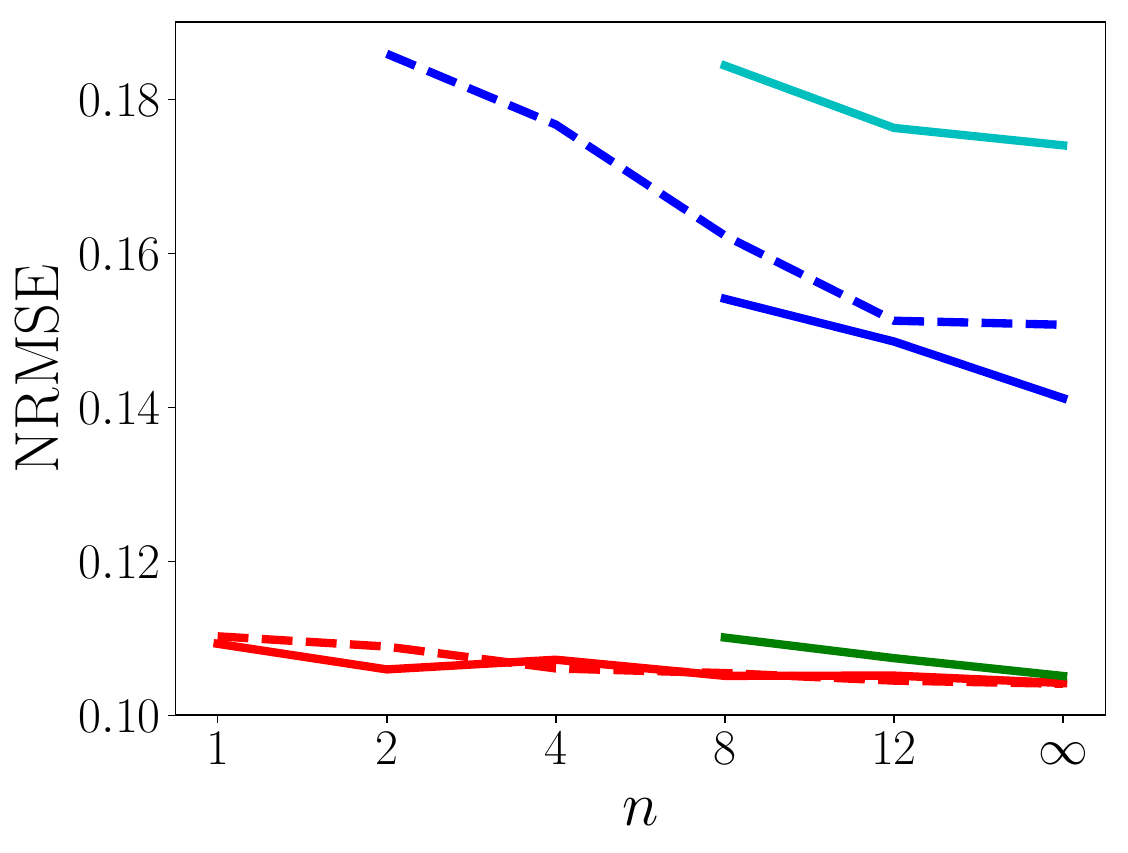}

\end{overpic}
\vspace{0.2cm}
 \begin{overpic}[width=0.49\linewidth,tics=10]{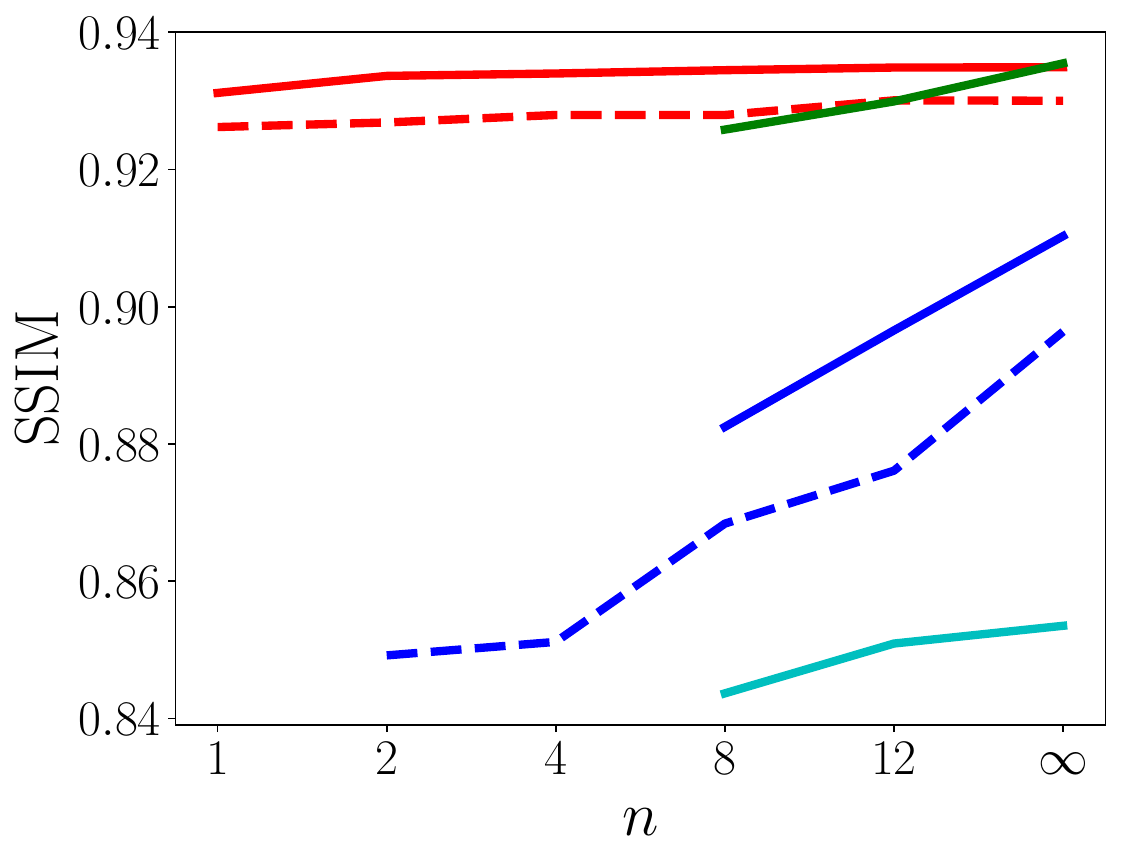}

\end{overpic}
 \begin{overpic}[width=0.49\linewidth,tics=10]{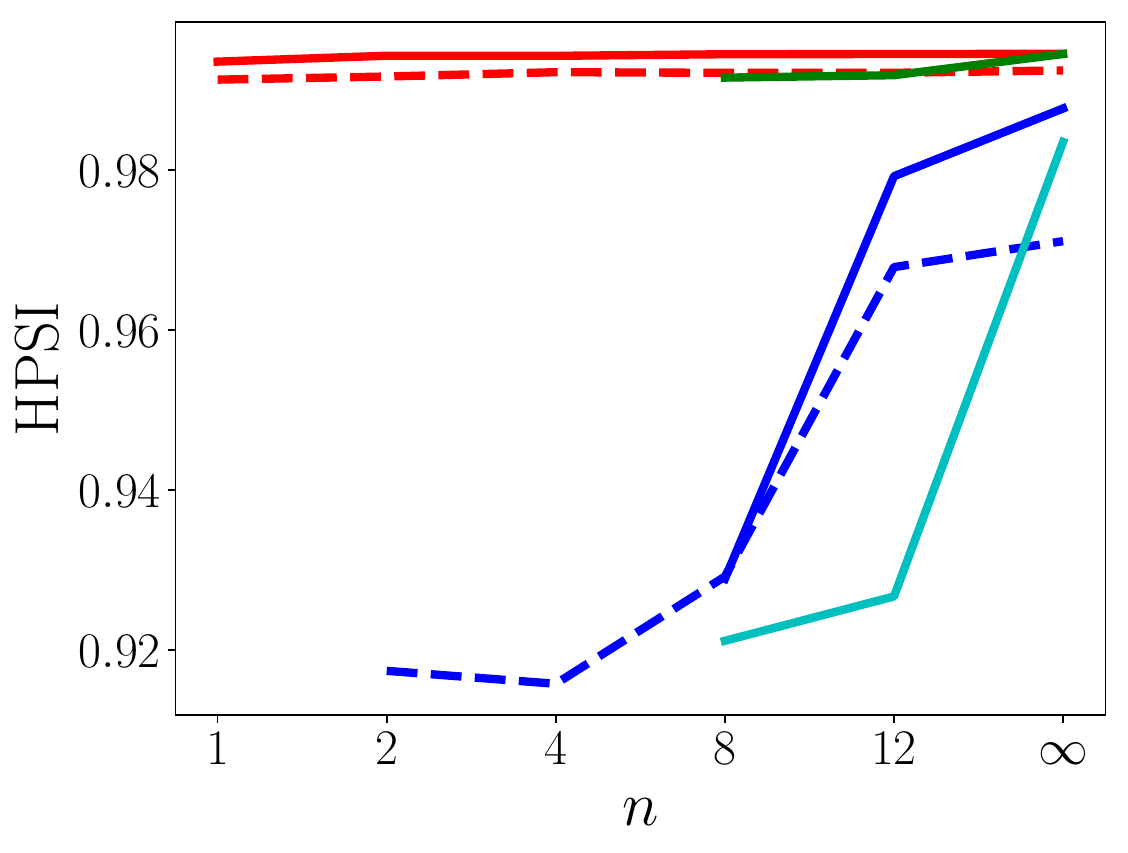}

\end{overpic}
 \begin{overpic}[width=\linewidth,tics=10]{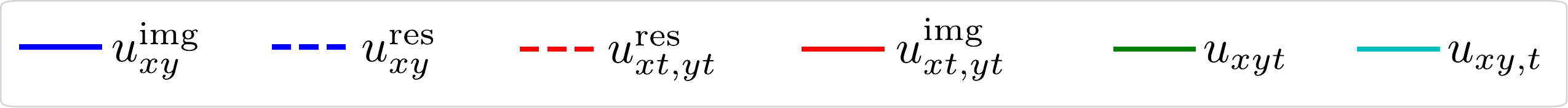}

\end{overpic}
\end{minipage}
\caption{Quantitative measures for all discussed Deep Learning-based post-processing methods when trained on datasets including different number of subjects $n$. Missing values for some $n$ denote that the network was not properly trainable on the restricted dataset.}\label{DL_nP_variation}
\end{figure}

The spatially trained U-net $u_{xy}^{\mathrm{img}}$ correctly removed the undersampling artefacts in the spatial domain. However, the reduction of the artefacts is less accurate than for $u_{xt,yt}^{\mathrm{img}}$, see Figure \ref{DL_comparison_fig} (b) and (e). Although we report a successful training in terms of consistent decrease of training as well as validation error, the model $u_{xy,t}$ poorly removed the artefacts. Intuitively, the temporal incoherence of the radial undersampling pattern which differs from the one in \cite{sandino2017deep} hinders the learning of the residual manifold and the network is therefore not suitable for our used undersampling scheme. Further, in \cite{sandino2017deep}, a zero-filled reconstruction is used as input of the network and therefore, the relation between the manifolds of the residuals and the ground truth images might differ as well from our case. In contrast, learning the manifold of ground truth sequences is highly facilitated by the temporal correlation of the $2D$ frames. In fact, already a network with one single convolutional layer with $N_t$ channels and $64$ filters accurately removed all the artefacts from the image sequence. However, temporal information is lost and we point out we were not able to obtain satisfactory results by the application of deeper networks. 
The $3D$ U-net $u_{xyt}$ and our proposed method $u_{xt,yt}^{\mathrm{img}}$ perform comparably well. Both correctly removed the undersampling artefacts in spatial as well in spatio-temporal domain and led to a good preservation of the heart movement. In terms of the image-error-based PSNR and NRMSE measures, our method performs slightly better than the $3D$ U-net $u_{xyt}$ which, on the other hand, yields slightly better results in terms of SSIM and HPSI. However, the differences are marginal and barely visible. Further, note how our proposed method achieves similar results as the $3D$ U-net $u_{xyt}$ even when trained on one single patient, see Table \ref{table_n_patients_variation}. Figure \ref{DL_nP_variation} shows the statistics calculated on the $2D$ frames for all different discussed Deep Learning-based post-processing approaches where the number of subjects $n$ contained in the training dataset was varied. The case $n=\infty$ corresponds to $n=12$ with all previously mentioned data-augmentation techniques. Clearly, our proposed method of training on the $2D$ spatio-temporal slices is the most suitable for obtaining satisfactory results when training a network on a highly limited dataset. The models $u_{xt,yt}^{\mathrm{img}}$ and $u_{xt,yt}^{\mathrm{res}}$ are the only ones to allow the successful training of a network on data extracted from one single subject. For $u_{xy}^{\mathrm{img}}$ and $u_{xy}^{\mathrm{res}}$, the results obtained for $n=2$ and $n=4$ were obtained by early stopping due to early overfitting. The models $u_{xy,t}$ and $u_{xyt}$ are properly trainable starting from $n=8$. The $3D$ U-net $u_{xyt}$ and our method $u_{xt,yt}^{\mathrm{img}}$ achieve comparable performance in terms of the reported measures for $n=\infty$.

\subsection{Comparison with State-of-the-Art Iterative Reconstruction Methods}

\begin{figure*}[!h]				
\begin{minipage}{\linewidth}
\centering
\begin{overpic}[width=\linewidth,tics=2]{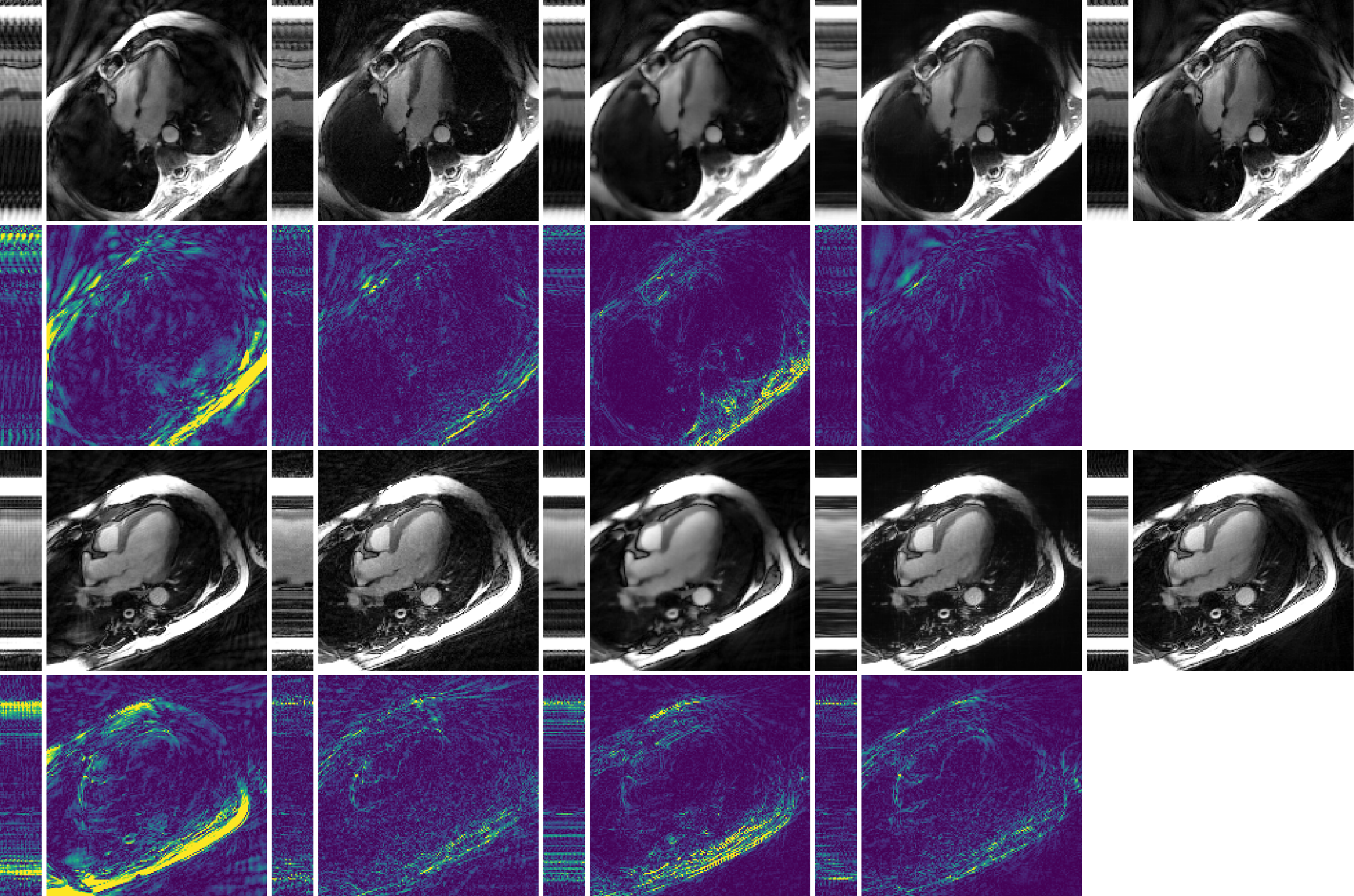}
\put (16,51) {\Large\textcolor{white}{(a)}}
\put (36,51) {\Large\textcolor{white}{(b)}}
\put (56,51) {\Large\textcolor{white}{(c)}}
\put (76,51) {\Large\textcolor{white}{(d)}}
\put (96,51) {\Large\textcolor{white}{(e)}}
\put (16,18) {\Large\textcolor{white}{(f)}}
\put (36,18) {\Large\textcolor{white}{(g)}}
\put (56,18) {\Large\textcolor{white}{(h)}}
\put (76,18) {\Large\textcolor{white}{(i)}}
\put (96,18) {\Large\textcolor{white}{(j)}}
\end{overpic}
\end{minipage}
\caption{Comparison with different state-of-the-art iterative reconstruction methods. $kt$-FOCUSS (a) and (f), TV$+$TVT (b) and (g), DL$+$TV (c) and (h), proposed method (d) and (i), $kt$-SENSE    reconstruction with $N_{\theta} = 3400$ radial lines. The point-wise error images are magnified by a factor of $\times 3$. All images are displayed on the same scale.}\label{IR_comparison_fig}
\end{figure*}
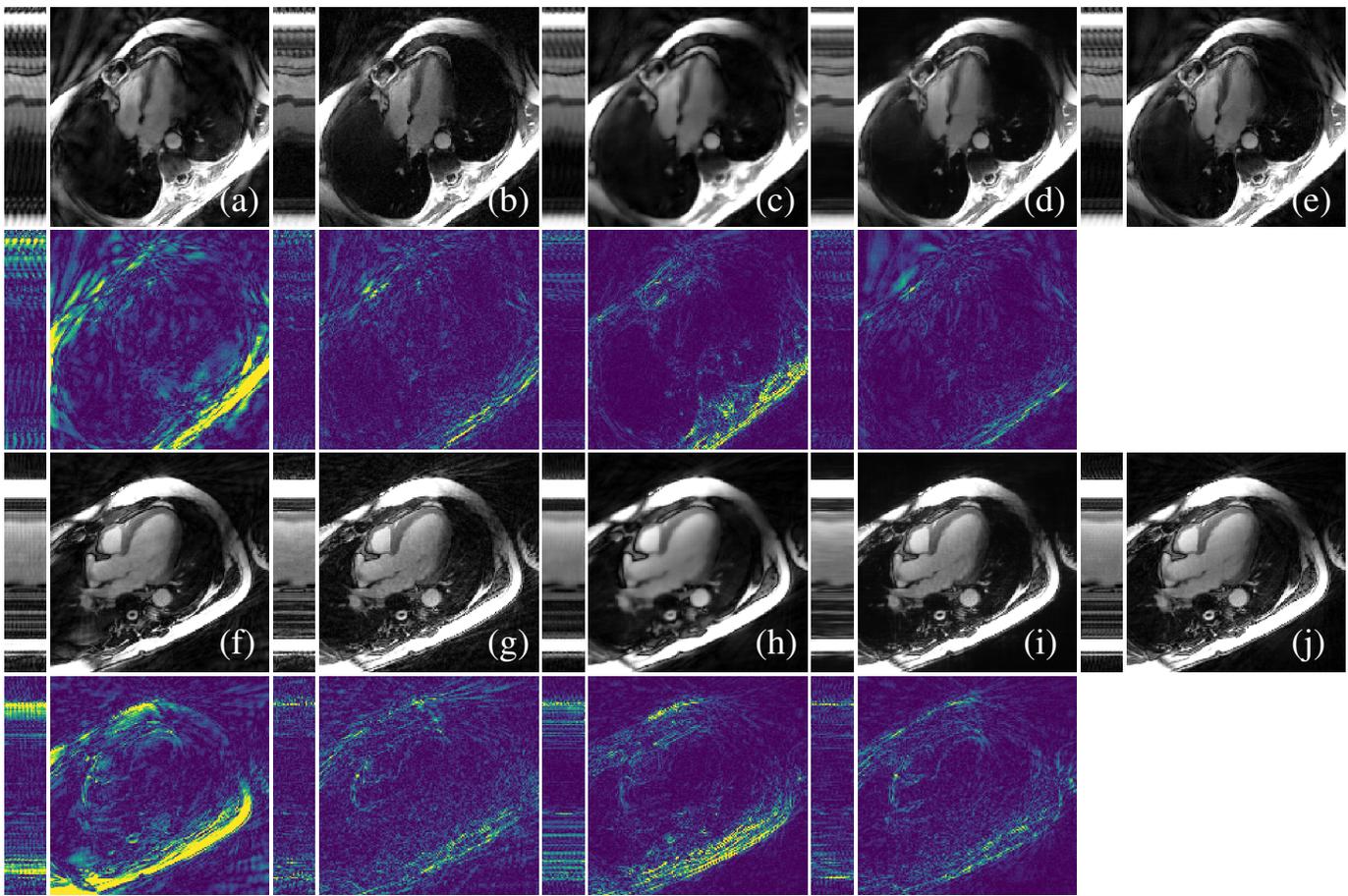

Here, we compare our proposed approach to established state-of-the-art iterative reconstruction methods for cine cardiac MRI. Since iterative reconstruction methods are time consuming, we only reconstructed images from the patients' data which corresponds to one training/validation/testing setting of our four-fold cross-validation set-up. 
For comparison, images were reconstructed with $kt$-FOCUSS, a CS-based approach \cite{Jung2009}, an iterative reconstruction approach using spatial and temporal total variation (TV$+$TVT) for regularization \cite{block2007undersampled} and a method employing regularization based on learned spatio-temporal  dictionaries as well as spatial and total variation minimization (DL$+$TV) \cite{wang2004image}. The latter method was extended by combining the approach proposed in \cite{wang2004image} with \cite{caballero2014dictionary} by learning the dictionaries jointly from the real and imaginary part of the image data. Further, we extended the method to be applicable to  multi-coil datasets. We implemented the method using the operator discretization library (ODL) \cite{adler_github} for all needed operators.\\ 
Figure \ref{IR_comparison_fig} shows examples of the results obtained on the patients' data for the mentioned iterative reconstruction methods and our proposed model $u_{xt,yt}^{\mathrm{img}}$. Although our method was trained on healthy volunteers, pathological heart wall motion (septal flash in Figure \ref{IR_comparison_fig} (a)-(e) and hypo-kinetic anterior and posterior wall with strongly reduced ejection fraction in Figure 13 (f) - (j)) is clearly visible with the proposed method. Also small features, such as the chordae tendinae connecting the valves and the papillary muscles, are well preserved, see Figure \ref{IR_comparison_fig} (i). 
Table \ref{IR_comparison_table} shows the obtained results with the iterative reconstruction methods as well as with our proposed network $u_{xt,yt}^{\mathrm{img}}$. We see that our method clearly outperforms the methods $kt$-FOCUSS and TV$+$TVT with respect to all reported quantitative measures. The most significant increase of performance is achieved against $kt$-FOCUSS, where, on the $2D$ frames, our method yields an increase of approximately $6$\,dB, $4.9 \%$ and $2\%$ in terms of PSNR, SSIM and HPSI. Further, our proposed method's NRMSE is approximately half of the one of  $kt$-FOCUSS. TV$+$TVT surpasses $kt$-FOCUSS in terms of all reported measures. Even if DL$+$TV surpasses TV$+$TVT with respect to all reported measures but HPSI, DL$+$TV tends to slightly smooth image details, possibly caused by a too strong regularization as well as the smoothing effect of the average of the reconstruction from patches. Further, note that the complex-valued patches were obtained by a disjoint sparse coding of the real and imaginary part of the patches as in \cite{caballero2014dictionary}.  Our method $u_{xt,yt}^{\mathrm{img}}$ outperforms DL$+$TV with respect to all reported measures except for SSIM on the spatio-temporal slices. Note that the reconstruction time for DL$+$TV is higher than for our method by several orders of magnitude, see Section \ref{reco_time_subsec}.
\begin{table}[!h]

\renewcommand{\arraystretch}{1.3}

 \caption{Comparison with different Iterative Reconstruction Methods.}\label{IR_comparison_table}
\centering

\begin{tabular}{l|S S S S}
   \hline
     \textbf{Reconstruction}  &  {\textbf{$kt$-FOCUSS}} & {\textbf{TV$+$TVT}}    &{\textbf{DL$+$TV}} & {\textbf{$u_{xt,yt}^{\mathrm{img}}$}}   \\
       \hline
       &  \multicolumn{4}{c}{\textbf{Statistics on $2D$ Frames}}\\
    \textbf{PSNR}  & 	31.2310529212 	&	34.7939244895 	&   35.1537944856 	&   37.5720774272	 	 	\\
     \textbf{SSIM} & 	0.887116866908	&  	0.916216972274 	&   0.932230680493	&   0.933310720507 		\\
     \textbf{HPSI} & 	0.965754644741	&   0.98728772695 	&   0.979519948768    &   0.992302064436  	 	 \\
    \textbf{NRMSE} & 	0.212870843533	& 	0.139954747509 	&   0.131606555002	&	0.102031499147    	\\
   \hline
     & \multicolumn{4}{c}{\textbf{Statistics on $2D$ Spatio-Temporal Slices}}\\

    \textbf{PSNR}  &	24.4934876485	&	27.0920903583 	&   27.9419870103	& 	29.5536561704      	 	\\
     \textbf{SSIM} &	0.735399705456	&   0.78569776835 	&   0.835996248094	& 	0.800250929794     	\\
     \textbf{HPSI} & 	0.970059721003	&  	0.988553940754	&  	0.978428190363 	& 	0.991612469309      	 \\
    \textbf{NRMSE} & 	0.25667045477 	&   0.203131690563	&   0.170685496252	&	0.157535939056      	\\
    \hline
  \end{tabular}
\end{table}

\subsection{Comparison with State-of-the-Art Cascaded Networks}

For the sake of completeness, we compare our method to the two state-of-the-art methods for $2D$ cine MRI based on cascaded networks presented in \cite{schlemper2017deep} and \cite{qin2019convolutional}. 
Cascaded networks combine iterative reconstruction methods and NNs in the sense that they  can be interpreted as unrolled iterative schemes where the networks play the role of regularizers learned from data  \cite{hammernik2018learning, kobler2017variational,  adler2018learned, adler2017solving}. 
While the NNs remove the artefacts from the undersampled image reconstructions, the data-consistency (DC) layers ensure that the outputs provided by the single networks match the measured data in $k$-space domain.
In \cite{schlemper2017deep}, the used NNs are $3D$ CNNs, while in \cite{qin2019convolutional}, the $3D$ CNNs are replaced by $2D$ recurrent CNNs. For the comparison, we used the codes available in \cite{schlemper2017deep} and \cite{qin2019convolutional}. Note that the main underlying assumption for cascaded networks is that the forward and adjoint operators  can be integrated in the network architecture.  For our data, the forward operator is given by a NUFFT encoding operator which measures $k$-space data from $n_c =12$ coils. Since building a deep cascade of CNNs is not possible by including our operator in the DC layers, we trained the networks on the image and $k$-space data for each coil separately. The final image estimates were then obtained by combining the images from the single coils using coil sensitivity information.
Table \ref{cascaded_CNNs_results} summarizes the results of the cascaded networks. The $3D$ CNN cascade approach yields slightly better image quality metrics compared to our approach, most probably due to the integration of the forward and adjoint operators in the DC layers. Note that for this experiment, the input images $\mathbf{x}_I$ were retrospectively simulated from the $kt$-SENSE reconstructions $\mathbf{x}$ and therefore, the statistics for our approach differ from the ones reported in  Tables IV and V, where the images are reconstructed from real $k$-space data obtained from the scanner.  Further, we report that, even if we did not observe overfitting, for the fold where the test set consists of patient data, the cascaded networks show a significant decrease in performance. This might indicate that the networks are more susceptible to possible significant differences between the training and test set data.
Figure \ref{DL_cascaded_CNNS_fig} shows qualitative results for the comparison of the two cascaded networks  and our approach. The statistics in Table \ref{cascaded_CNNs_results} were obtained by averaging the results on the test set for each fold. On each test set, the measures were obtained by testing the networks for which the trainable parameters led to the smallest average error on the whole validation set. The results for the different folds can be found in the supplementary materials which are available in the multimedia tab.
\begin{figure}[!t]				
\begin{minipage}{\linewidth}
\centering
\begin{overpic}[width=\linewidth,tics=3]{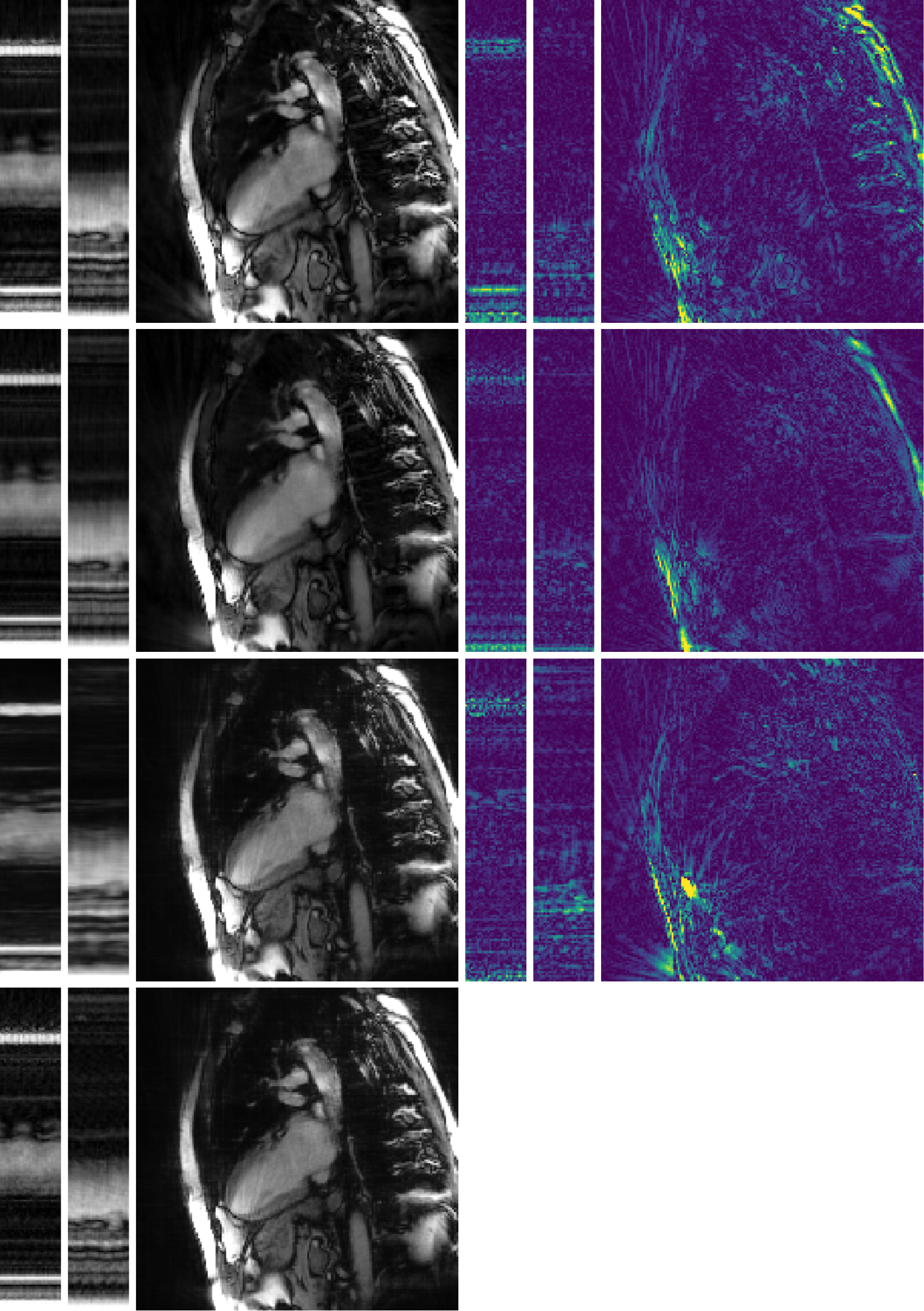}
\put (29,77) {\Large\textcolor{white}{(a)}}
\put (29,52) {\Large\textcolor{white}{(b)}}
\put (29,27) {\Large\textcolor{white}{(c)}}
\put (29,2) {\Large\textcolor{white}{(d)}}
\end{overpic}
\end{minipage}
\caption{Comparison with different cascaded CNNs: $2D$ CRNN Cascade (a), $3D$ CNN-Cascade (b), proposed (c) and the reference $kt$-SENSE reconstruction (d). The Figure show results for the fold where only patient's data is included in the test set. Qualitatively, all the three methods perform similarly.}\label{DL_cascaded_CNNS_fig}
\end{figure}
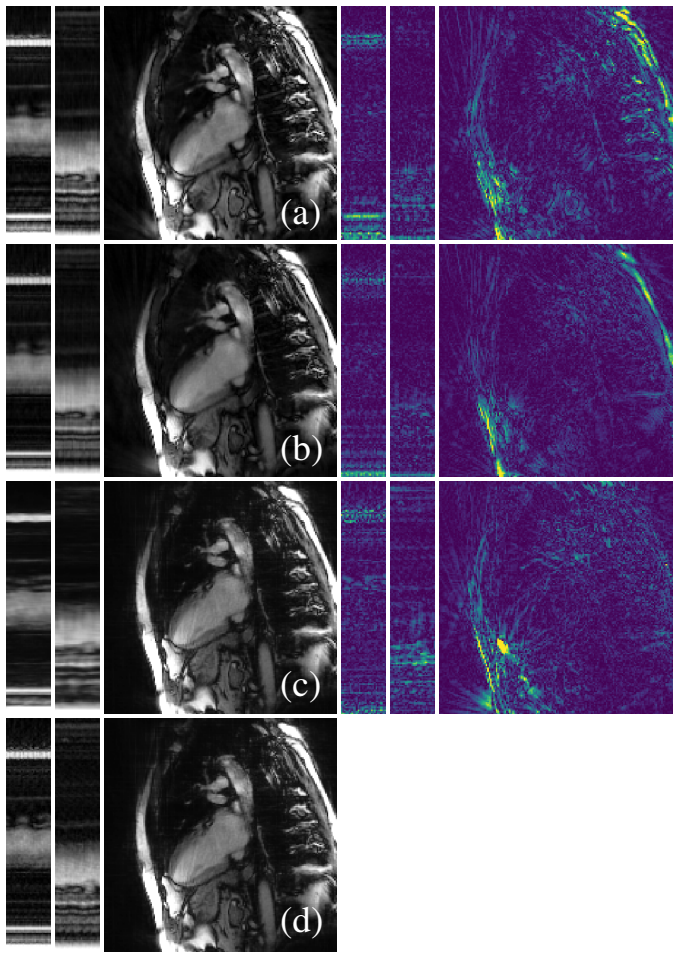
\begin{table}[!h]

\renewcommand{\arraystretch}{1.3}

 \caption{Comparison with different cascaded CNNs.}\label{cascaded_CNNs_results}
\centering

\begin{tabular}{l|S S S}
   \hline
     \textbf{Reconstruction}  &  {\textbf{$3D$ CNN cascade}} & {\textbf{$2D$ CRNN cascade}}  & {\textbf{$u_{xt,yt}^{\mathrm{img}}$}}   \\
       \hline
       &  \multicolumn{3}{c}{\textbf{Statistics on $2D$ Frames}}\\

    \textbf{PSNR}  & 	41.83056542374743 & 37.94460160156524 	&   40.37607253124864	 	 	\\
     \textbf{SSIM} & 	0.9685580024295763 & 0.9604203288869013 	&   0.9538056020311377 		\\
     \textbf{HPSI} & 	0.9886794614408394 & 0.9725174663177436 	&   0.9893522092270026  	 	 \\
    \textbf{NRMSE} & 	0.06756184392024697 & 0.10275897197425365	&	0.07924168524686906    	\\
   \hline
     & \multicolumn{3}{c}{\textbf{Statistics on $2D$ Spatio-Temporal Slices}}\\

    \textbf{PSNR}  &	33.77854005840598 & 30.38266345942381 	& 	32.28072916070096    	 	\\
     \textbf{SSIM} &	0.9075167389780251 & 0.8853252458824772 	& 	0.8419932192851152    	\\
     \textbf{HPSI} & 	0.9879646832277409 & 0.970050222137147 	& 	0.9848878794293197     	 \\
    \textbf{NRMSE} & 	0.10399272640167893 & 0.1401736494583143	&	0.1264952402561903     	\\
    \hline
  \end{tabular}
\end{table}
\subsection{Reconstruction Times}\label{reco_time_subsec}
We report the reconstruction times needed for the reconstruction of the images with the different previously discussed methods. First, we note that the methods employing iterative reconstruction are the most demanding in terms of computational times. $kt$-FOCUSS, $kt$-SENSE and TV$+$TVT are in the same range, where the reconstruction times per slice vary from approximately 110\,s to approximately 180\,s. The DL$+$TV method is by far the most computationally  expensive method, as the regularized inverse problem has to be solved for each coil separately. Therefore, the average overall reconstruction time per slice amounts to roughly  13\,000\,s, where nearly 1\,500\,s are needed by ITKrM \cite{schnass2016convergence} which replaced the computationally heavier $K$-SVD \cite{aharon2006k}, 7\,800\,s by the sparse coding with orthogonal matching pursuit, 310\,s for the reconstruction from the sparsely approximated patches and 2\,058\,s for the preconditioned conjugate gradient (PCG) method.\\
Note that we trained all the $2D$ U-nets on image sequences which were previously cropped to $220 \times 220 \times 30$. Also, due to memory limits, the shape of the image sequences which are processed by the  $3D$ U-net was $128 \times 128 \times 20$. Therefore, for the methods $u_{xy}$,  $u_{xy,t}$ and $u_{xyt}$, the $320\times 320 \times 30$ image-sequences were reconstructed  from patches. In particular, we used strides of size $25 \times 25 $ for the spatial and spatio-temporal $2D$ U-nets and strides of $32 \times 32 \times 5$  for the $3D$ U-net, resulting in $5 \cdot 5 \cdot 30 = 750$, $ 5 \cdot 5 = 25$ and $7 \cdot 7 \cdot 3 = 147$ samples to be processed for the reconstruction of a single slice. For our method $u_{xt,yt}^{\mathrm{img}}$, the strides are $50$ (in $x$- and $y$ direction), resulting in $3 \cdot (220 + 220)$ samples to be processed per slice. Processing one sample on a Titan Xp GPU takes on average $0.0093$\,s for $u_{xy}^{\mathrm{img}}$ and $u_{xy}^{\mathrm{res}}$, $0.0236$\,s for $u_{xy,t}$, $0.0340$\,s for $u_{xyt}$ and $0.0034$\,s for our proposed approaches $u_{xt,yt}^{\mathrm{img}}$ and $u_{xt,yt}^{\mathrm{res}}$. Table \ref{reco_times} summarizes the reconstruction times for a slice of size $320 \times 320 \times 30$ for all the reported methods with the  aforementioned strides. The times needed to denoise a slice obviously heavily depend on the number of patches the sequence is reconstructed from and could be easily  reduced by using larger strides. For the $2D$ methods, one could also obtain the $320 \times 320 \times 30$ image sequences by directly applying the networks to the $320 \times 320 \times 30$ samples. Note that for the $3D$ U-net this not possible because of memory limits. 
The training times needed for the $2D$ CRNN cascade and the $3D$ CNN cascade amounted to approximately 1 day and 3 days and 14 hours while processing a single slice and all cardiac phases takes about 16.8\,s and 8.8\,s, respectively,. Note that the reconstruction of one slice involves the processing of the images of all $n_c=12$ coils.

\begin{table}[!h]

\renewcommand{\arraystretch}{1.3}

 \caption{Comparison of the Reconstruction Times Slice}\label{reco_times}
\centering

\begin{tabular}{lc}
   \hline
   	{\textbf{Method}} &  {\textbf{Reconstruction  Time $[s]$}} \\
   	\hline
   	   {\textbf{NUFFT }} & 5 \\
       {\textbf{NUFFT $+ \ u_{xy}^{\mathrm{res}} / u_{xy}^{\mathrm{img}}$ }}  & $5+7$   \\ 
       {\textbf{NUFFT $+ \ u_{xy,t}$}} & $5 + 0.64$ \\
       {\textbf{NUFFT $+ \ u_{xyt}$ }}  & $5 +5$   \\
       {\textbf{NUFFT $+ \ u_{xt,yt}^{\mathrm{res}} / u_{xt,yt}^{\mathrm{img}}$  }} & $5+4.4$  \\
       {\textbf{$kt$-FOCUSS}} & 110  \\
       {\textbf{$kt$-SENSE}} & 150 \\
       {\textbf{TV$+$TVT}} & 180 \\
       {\textbf{DL$+$TV}} & 13\,036 \\
       {\textbf{$2D$ CRNN cascade}} & 16.8 \\
       {\textbf{$3D$ CNN cascade}} & 8.8 \\
    \hline
  \end{tabular}
\end{table}

\section{Discussion and Conclusion}\label{section_Discussion_and_Conclusion}

In this work, we have presented a new approach for the task of undersampling artefacts reduction in $2D$ cine MRI. Even if the employed U-net is a widely used network architecture for various inverse problems, to the best of our knowledge, this is the first work in which the U-net is applied to $2D$ spatio-temporal slices. We have investigated and demonstrated several advantages of the approach compared to the training in the spatial domain. Consistent with \cite{han2016deep, bae2017beyond, LeeDeep2017}, the performed persistent homology analysis confirms the motivation that the superiority of the proposed approach can be attributed to the simpler topological complexity of the two-dimensional spatio-temporal slices. Further, the analysis suggests that the architecture should be chosen such that the network is trained to learn the ground truth images rather than the residuals. Note that our analysis is consistent with the results presented in \cite{jin2017deep} and \cite{han2016deep}, where streaking artefacts resulting from a sparse view CT acquisition are most efficiently removed when U-net learns the residual manifold, which was shown to have a lower complexity than the one of the ground truth images \cite{han2016deep}. This is  related to the fact that the undersampling pattern in sparse view CT is regular. Conversely, in CS MRI, where the undersampling schemes, e.g. golden-angle radial undersampling, are designed to be incoherent with the assumed sparsifying basis \cite{lustig2008compressed}, one would expect the residual manifolds to have a more complex topological structure and therefore, the network's architecture should be chosen appropriately. Further investigation of the relation between the topological complexity of the residuals and the artefact-free images in different imaging modalities and the performance of the trained networks will be investigated in the future.

Our approach allows to successfully train a U-net on highly limited data, overcoming the problem of unavailability of large datasets or the need to rely on data-augmentation. We demonstrated that our method already outperforms the spatially trained U-net when trained on one single healthy volunteer in terms of all quantitative measures. When trained on a small number of  volunteers, our network is already able to accurately preserve the heart movement and delivers results which are similar to the ones obtained when training on 12 subjects. In contrast to the spatial training approach, the proposed method naturally almost achieves rotation-equivariance by the sole change of perspective on the data. The network does therefore neither require changes in the architecture, nor data-augmentation based on rotation to achieve this property. Clearly, the reason lies in how a rotation in image space results in a transformation similar to a translation in the spatio-temporal domain, and therefore, since the network consists of convolutional and max-pooling layers, it is stable with respect to rotation in image space. 
Even if the reconstruction of a single slice and all its cardiac phases requires the evaluation of a large number of samples, reconstruction is fast and can be achieved in approximately  $4.4$\,s on a Titan Xp GPU.\\
As discussed in \cite{ye2018deep} and \cite{han2018framing}, the U-net tends to smooth out image details when trained in the spatial domain. In the proposed approach, however, image details in the spatial domain are well preserved. Our method, on the other hand, well preserves image details and further outperforms all other tested $2D$ CNNs with respect to all reported measures and achieves results comparable to the $3D$ U-net even when trained only on two subjects. 
Due to the small size of the data when considered in spatio-temporal domain, training times could be shortened to 3 hours compared to 6 hours for the $3D$ U-net. Further, since the spatio-temporal manifold $\mathcal{M}_{xt,yt}^{\mathrm{img}}$ has a particularly simple structure, the reducing the artefacts reduces to a simpler task than in the spatial domain and training times could be further reduced by earlier stopping the training.

As for all Deep Learning-based post-processing methods, the main limitation of our proposed method is the possible lack of data-consistency. Even if our method is based on post-processing of the magnitude images, the method could be easily extended to process the real and imaginary part of the spatio-temporal slices separately. Therefore, handling complex-valued data does not represent a limitation and data-consistency could be enforced by for example performing several iterations of PCG for minimizing a properly chosen functional including a data-consistency and regularization term based on the output of our method, see for example \cite{wang2016accelerating}.

We have compared our proposed method to several state-of-the-art methods for iterative reconstruction in dynamic MRI. Our method outperforms $kt$-FOCUSS and TV$+$TVT with respect to all reported measures and achieves similar results as the dictionary learning- and total variation-based method DL$+$TV. However, our method is faster than DL$+$TV by several orders of magnitude as it performs a one-step regularization based on an initial NUFFT reconstruction. 
The iterative reconstruction methods $kt$-FOCUSS, TV$+$TVT and DL$+$TV used for comparison require the tuning of several parameters which were kept fixed for all patients. Therefore, further patient-specific parameter tuning might further improve the image quality in Figure \ref{IR_comparison_fig} (a), (b), (c), (f), (g) and (h). In particular,  DL$+$TV makes specific parameter tuning difficult due to its prohibitive reconstruction times.

Further, we have compared our method with two state-of-the-art methods based on cascaded CNNs \cite{schlemper2017deep, qin2019convolutional} trained on retrospectively simulated data.  Although the $3D$ cascaded network's performance is slightly superior to our method, note that for the cascades the input images are zero-filled reconstructions using a Cartesian mask whose support is given by the indices of the $k$-space coefficients which were interpolated from the radially acquired $k$-space data. Therefore, the input images for the cascades contain artefacts which are inherently different from the ones obtained by our NUFFT reconstruction using $n_c=12$ coils and $N_{\theta}=1130$ spokes. Also, even if our method only performs subsequent post-processing, the obtained results are qualitatively competitive with the ones obtained by the cascaded networks and we point out that our approach could also be easily extended to be integrated in cascaded networks. This will be subject of future work.

In this work, we used $kt$-SENSE to obtain the ground truth samples from a 10\,s breathhold. Although this yielded high image quality, residual undersampling artefacts which might impair the trained U-net might still be visible. Also, $kt$-SENSE  already makes assumptions about the temporal smoothness of the image data. Therefore, further improvement of our method might be achieved by increasing the duration of the breathhold scan to achieve higher ground truth-image quality. 
\section*{Acknowledgements}
A. Kofler and M. Dewey acknowledge the support of the German Research Foundation (DFG), project number GRK 2260, BIOQIC. C. Wald and M. Dewey have received funding from the DHA of the Berlin Institute of Health.
\bibliographystyle{IEEEtran}
\bibliography{IEEEabrv,spatio_temporal_unet}

\end{document}